\documentclass[traditabstract]{aa}
\usepackage{txfonts}
\usepackage{graphicx}
\usepackage{natbib}
\bibpunct{(}{)}{;}{a}{}{,} 

\begin{document}

\title{Cross-calibration of Suzaku XIS and XMM-Newton EPIC using clusters of galaxies}

\author{K.~Kettula\inst{\ref{inst1}}\and J.~Nevalainen\inst{\ref{inst1},~\ref{inst2},~\ref{inst3}} \and E.D.~Miller\inst{\ref{inst4}}}

\institute{Department of Physics, Dynamicum, PO Box 48, 00014, University of Helsinki, Finland \\ 
\email{kimmo.kettula@iki.fi}\label{inst1}
\and
Finnish Centre for Astronomy with ESO, University of Turku, V\"ais\"al\"antie 20, 21500 Piikki\"o, Finland \\ 
\email{jukka.h.nevalainen@helsinki.fi}\label{inst2}
\and
Tartu Observatory, 61602 T\~oravere, Estonia \label{inst3}
\and
Kavli Institute for Astrophysics and Space Research, Massachusetts
Institute of Technology, 77 Massachusetts Avenue, Cambridge, MA 02139, USA\label{inst4}}

\date{Received / Accepted}

\abstract{We extend a previous cross-calibration study by the International Astronomical Consortium for High Energy Calibration 
(IACHEC) on XMM-Newton/EPIC, Chandra/ACIS and BeppoSAX/MECS X-ray instruments with clusters of galaxies 
(Nevalainen et al., 2010) to Suzaku/XIS instruments. Our aim is to study the accuracy of the energy-dependent effective area
calibration of the XIS instruments using observations during 2005--2008. 
By comparison of the spectroscopic  temperatures, fluxes and fit residuals obtained with Suzaku/XIS and XMM-Newton/EPIC-pn
 for the same cluster, we evaluated the systematic uncertainties of the energy dependence 
 and the normalisation of the effective area between different detectors.
The temperatures measured in the hard 2.0--7.0 keV energy band with all instruments are consistent within $\sim$5 \%.
However, temperatures obtained with the XIS instruments in the soft 0.5--2.0 keV band disagree by 9--29 
\%. We investigated residuals in the XIS soft band, which showed that if XIS0 effective area shape is accurately 
calibrated, the effective areas of XIS1 and XIS3 are overestimated below $\sim$1.0 keV (or vice versa). 
Adjustments to the modelling of the column density of the XIS contaminant in the 3--6 arcmin extraction region 
while forcing consistent emission models in each instrument for a given cluster significantly improved the fits. 
The oxygen column density in XIS1 and XIS3 contaminant must be increased by $\sim$1--2 $\times$ 10$^{17}$ cm$^{-2}$ in 
comparison to the values implemented in the current calibration,while the column density of the XIS0 contaminant given 
by the analysis is consistent with the public calibration.  XIS soft band temperatures obtained with the modification to 
the column density of the contaminant agree better with temperatures  obtained with the EPIC-pn instrument of XMM-Newton, 
than with those derived using the Chandra-ACIS instrument.
However, comparison of hard band fluxes obtained using Suzaku-XIS to fluxes obtained using the Chandra-ACIS and EPIC-pn instruments 
proved inconclusive.
}


\titlerunning{Cross-calibration of Suzaku XIS and XMM-Newton EPIC using clusters of galaxies}
\maketitle

\section{Introduction}

Clusters of galaxies are bright, stable and extended objects and thus suitable for X-ray calibration. The X-ray spectra of clusters 
consist typically of a 
combination of thermal bremsstrahlung and collisionally excited line emission. Temperature measurements of clusters above kT $\sim$ 2 keV  are 
dominated by the shape of the X-ray continuum. Consequently, by comparing temperatures of the same cluster measured with different instruments, one can get 
an indication on the cross-calibration status of the energy dependence of the effective area (defined here as the product of the mirror effective area, 
detector quantum efficiency and filter transmission) between the instruments. Similarly the normalisation of the effective area affects the  
determination emission measure and cluster flux, and comparison of cluster fluxes measured with different instruments thus yields information about the 
relative normalisation of the effective areas of the instruments.  Analysis of the residuals of the spectral fits yields more detailed 
information on the cross-calibration.

The International Astronomical Consortium for High Energy Calibration (IACHEC)\footnote{http://web.mit.edu/iachec/}  was formed to provide 
standards for high energy calibration and to supervise cross-calibration between different observatories. This paper is based on the 
activities of the IACHEC clusters working group. The working group performed a study of clusters of galaxies in order to investigate the 
effective area cross-calibration status between XMM-Newton / EPIC, Chandra / ACIS and BeppoSAX / MECS instruments \citep{nevalainen10}. 
They found that while the energy dependence of the effective area is accurately calibrated in the hard energy band (2.0 -- 7.0 keV), 
the effective area normalisations  disagree by 5 -- 10 \%. They also found significant cross-calibration differences in the soft 
energy band (0.5 -- 2.0 keV). They determined that temperatures measured with Chandra / ACIS are systematically $\sim$ 20 \% higher than 
those measured with XMM-Newton / EPIC instruments. 
This amounts to a $\sim$ 10 \%  uncertainty of the ACIS and EPIC soft band cross-calibration of the effective area 
energy dependence. Due to the high statistical weight of the soft band data, \citet{nevalainen10} estimated that Chandra / ACIS 
and XMM-Newton / EPIC cluster temperatures measured in a typical 0.5 -- 7.0 keV energy band suffer from a systematic uncertainty of 
10 -- 15 \% due to the cross-calibration uncertainties.   

We extend the IACHEC cluster work to include Suzaku / XIS \citep{koyama07}. 	
We will evaluate the cross-calibration accuracy of 
the XIS instruments by comparison of measured temperatures and spectral fit residuals.
By comparing the measurements with XIS and EPIC-pn \citep{struder01} we aim to improve the knowledge about  
the soft band cross-calibration uncertainty between EPIC and ACIS.  \footnote{Due to the limited FOV of ACIS-S we cannot perform a one-to-one
comparison between the ACIS measurements and XIS.}

This paper is organised as follows.  In Section 2 we present the cluster sample used for our study, and in Section 3 we discuss the
selected spectral extraction regions.  We present data reduction and background modelling methods in Sections 4 and 5, respectively.  We
discuss the spectral analysis method in Section 6 and present the results from the analysis in Sections 7 and 8.  
In Section 9, we
investigate if uncertainties in the calibration of the optical blocking filter (OBF) molecular contaminant may affect the
cross-calibration of the effective area energy dependence.  Finally, we present our concluding remarks in Section 10. Throughout this paper,
we use 68 \% (1$\sigma$) confidence level for errors, unless stated otherwise.

\section{Cluster sample}
\label{sect:sample}

We constructed a cluster sample consists of HIFLUGS clusters \citep{HIFLUGCS} with XMM-Newton and Suzaku observations publicly available in June 2011, 
using our selection criteria detailed below. The clusters passing our selection criteria (see below) have been observed during the years 2005 -- 2008 
(Suzaku) and 2000 -- 2009 (XMM-Newton). The chosen clusters are all bright (flux $\sim 10^{-12}$ -- $10^{-10}$ erg / cm$^2$ / s in a 2 -- 7 keV band) and nearby 
($z \leq 0.0753$). General properties of the cluster sample are presented in Table \ref{tab:sample_prop} and  \ref{tab:obs_prop}.

\begin{table*}
\caption{General information on the cluster sample} %
\label{tab:sample_prop} 
\centering
\begin{tabular}{l c c c c}
\hline\hline
Name & N$_{\rm H}$ \tablefootmark{a} & $z$ \tablefootmark{b} & RA (J2000) \tablefootmark{c}  & Dec (J2000) \tablefootmark{c} \\ 
     & [$10^{20}$ cm$^{-2}$] &   & hh mm ss & \degr ~ \arcmin ~ \arcsec \\
\hline 
A1060      & 4.90 & 0.0126 & 10 36 42.6 & -27 31 44.2\\
A1795      & 1.19 & 0.0625 & 13 48 53.0 & 26 35 25.0 \\
A262       & 5.67 & 0.0163 & 01 52 46.0	& 36 09 09.1 \\
A3112      & 1.33 & 0.0753 & 03 17 57.7 & -44 14 18.3\\
A496       & 3.76 & 0.0329 & 04 33 38.1	& -13 15 40.0\\
AWM7       & 8.89 & 0.0172 & 02 54 27.4	&  41 34 46.8\\
Centaurus  & 8.30 & 0.0114 & 12 48 48.9 & -41 18 46.9\\
Coma       & 0.87 & 0.0231 & 12 59 35.7	&  27 57 34.0\\
Ophiuchus  & 19.3 & 0.0280 & 17 12 27.9 & -23 22 08.2\\
Triangulum & 11.5 & 0.0510 & 16:38:18.6 & -64:20:37.8\\
\hline 
\end{tabular}
\tablefoot{\tablefoottext{a}{LAB weighted average \citep{kalberla05}.}
\tablefoottext{b}{From NASA Extragalactic Database (http://ned.ipac.caltech.edu/)}
\tablefoottext{c}{Adopted coordinates of the cluster centre.}
}
\end{table*}

Our sample contains only such pointings where the off-axis angle between the cluster centre and the FOV centre is smaller 
than 1 arcmin. This ensures that we are studying the same detector regions and thus do
not fold in any additional calibration uncertainties that depend on the position of the detector.

We performed spectral analysis in two energy bands: 0.5 -- 2.0 keV (soft band) and 2.0 -- 7.0 keV (hard band).  We required that the observations 
have a minimum of 5 000 and a maximum of 500 000 data counts in the soft and hard energy bands.  The lower limit ensures sufficient 
number of counts to obtain reliable statistics, while the upper limit eliminates such high signal-to-noise data where other calibration uncertainties,
like gain and energy redistribution, are significant. We further excluded clusters where the total sky and non X-ray background exceeded 10 \% of 
the cluster flux (see Section \ref{sect:bkg} for details on background modelling).  

In order to minimise additional uncertainties in the soft energy band, we excluded clusters with a galactic absorption 
column density N$_{\rm H}$  $> 6 \times 10^{20}$ cm$^{-2}$ from the soft band sample (see Table \ref{tab:obs}).  
Thus the soft band sample contains a smaller number of objects than the hard band sample.

\section{Extraction regions}
\label{sect:regions}

\begin{figure*}
\centering
\includegraphics[width=17cm]{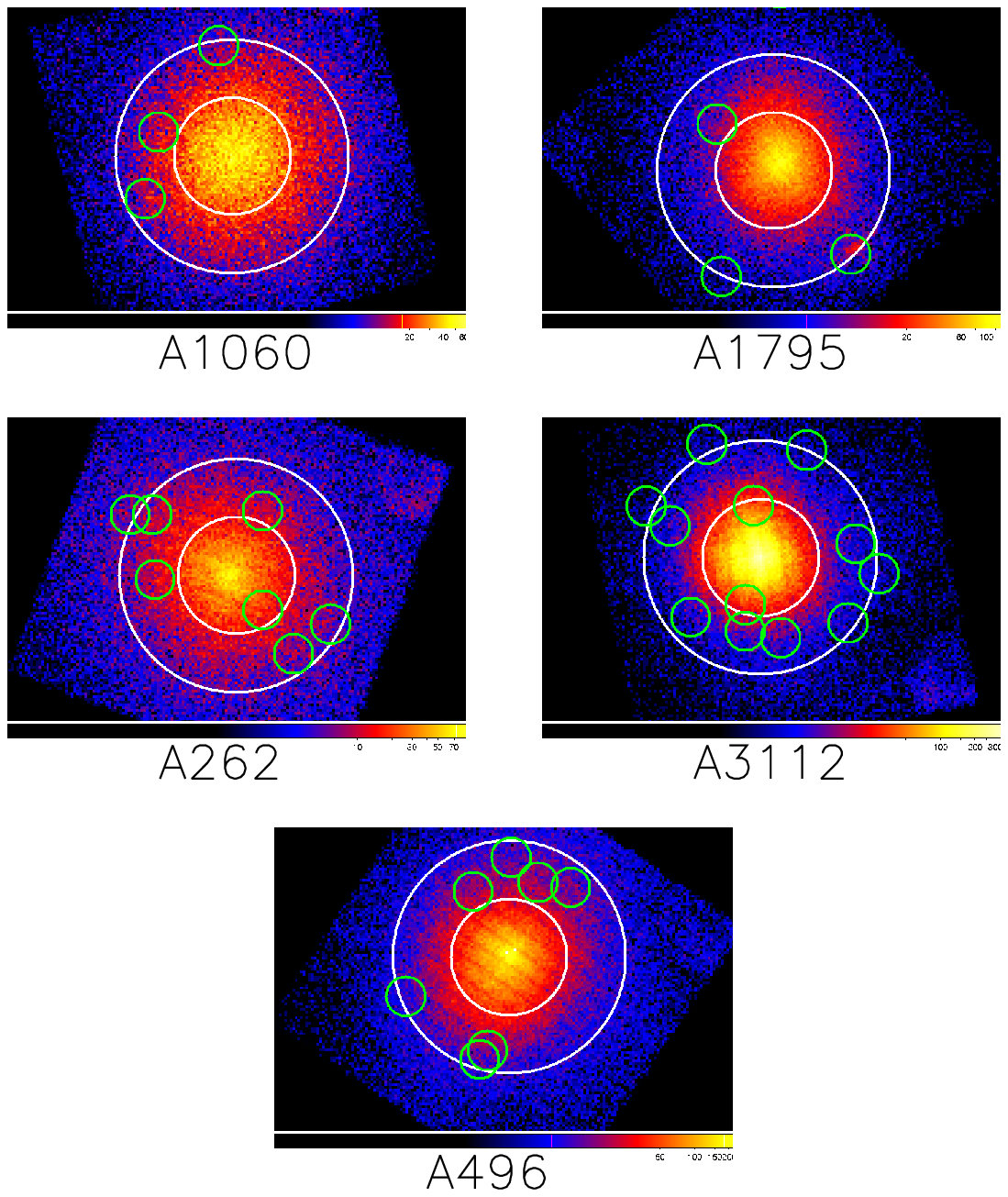}
\caption{XIS1 raw count rate maps (not corrected for the background or vignetting) of the soft band cluster sample in a 0.5 -- 7.0 keV energy band binned to 
8 pixels ($\sim$ 0.14 arcmin) per bin. The inner and outer radii of the 3 -- 6 arcmin extraction region 
are denoted by white circles, the green circles show the excluded regions centred on point sources seen in XMM-Newton EPIC / pn images. Scaling
of the maps are adjusted for plot clarity and do not reflect the relative brightness of the clusters. The colour bar shows the the total number of counts 
per bin.
}
\label{fig:xis1_reg}
\end{figure*}

As we attempted to minimise PSF scatter and background contamination, we chose to use 3 -- 6 arcmin annuli around the cluster centres as 
extraction regions (see Fig. \ref{fig:xis1_reg}). By excluding the inner 3 arcmin we minimise the Suzaku PSF scatter from the cool core 
(see Section \ref{sect:PSF}). The width of the annulus is larger than 
the Suzaku PSF so that the PSF scatter to and from the extraction region is minimised. By excluding the regions outside 6 arcmin distance from 
the cluster centre, i.e. maintaining in the bright central regions, we minimize the effect of background contamination.

The useful detector area (when excluding the CCD gaps and the regions around the calibration sources) varies somewhat between different XIS
instruments. The fraction of the full extraction annuli covered by the useful detector area varies at most by  12 \% between the different instruments. 
We thus assume that the temperature measurements are not significantly affected due to the above difference in the covered regions.

\section{Data reduction}
 
The general properties of the observations are listed in Table \ref{tab:obs_prop}. 

\begin{table*}
\caption{General information on the observations of the sample} 
\label{tab:obs_prop} 
\centering 
\begin{tabular}{l|c c c c|c c c c c} 
\hline\hline 
     & \multicolumn{4}{c|}{XMM-Newton} & \multicolumn{5}{c}{Suzaku} \\
Name & ID & Obs.date &  Exposure time (ks) \tablefootmark{a}  & pn mode \tablefootmark{b} & ID & Obs.date &\multicolumn{3}{c}{Exposure time (ks) 
\tablefootmark{a}} \\
     &  & yyyy-mm-dd & pn  &  &  & yyyy-mm-dd       & XIS0 & XIS1 & XIS3 \\
\hline
A1060       & 0206230101 & 2004-06-29 & 14 & F & 800003010 & 2005-11-22 & 30 & 30 & 30 \\ 
A1795       & 0097820101 & 2000-06-26 & 34 & F & 800012010 & 2005-12-10 & 10 & 10 & 10 \\
A262        & 0109980101 & 2001-01-16 & 16 & E & 802001010 & 2007-08-17 & 33 & 33 & 33 \\
A3112       & 0603050101 & 2009-07-21 & 63 & E & 803054010 & 2008-05-23 & 52 & 52 & 52 \\
A496        & 0506260401 & 2008-02-18 & 33 & F & 803073010 & 2008-08-02 & 31 & 31 & 31 \\
AWM7        & 0605540101 & 2009-18-16 & 76 & E & 801035010 & 2006-08-07 & 17 & 17 & 17 \\
Centaurus   & 0406200101 & 2006-07-24 & 52 & E & 800014010 & 2005-12-27 & 30 & 30 & 30 \\
Coma        & 0153750101 & 2001-12-04 & 17 & F & 801097010 & 2006-05-31 & 166 & 166 & 166 \\
Ophiuchus   & 0505150101 & 2007-09-02 & 10 & E & 802046010 & 2007-09-24 & 75 & 75 & 75 \\
Triangulum  & 0093620101 & 2001-02-17 & 3  & E & 803028010 & 2008-10-11 & 64 & 64 & 64 \\
\hline 
\end{tabular}
\tablefoot{\tablefoottext{a}{Clean exposure times after filtering, ~}\tablefoottext{b}{F = Full Frame, E = Extended Full Frame.}
}
\end{table*}

\subsection{Suzaku}

We used HEASoft\footnote{http://heasarc.nasa.gov/lheasoft/} release 6.11, containing Suzaku ftools version 18, for the processing. XIS calibration 
information was from XIS CALDB release 20110608 
\footnote{http://heasarc.gsfc.nasa.gov/docs/heasarc/caldb/data/suzaku/xis/ \\ 
index/cif\_suzaku\_xis\_20110608.html} 
and XRT calibration information from XRT CALDB release 20080709.
\footnote{http://heasarc.gsfc.nasa.gov/docs/heasarc/caldb/data/suzaku/xrt/ \\ 
index/cif\_suzaku\_xrt\_20080709.html}

We used the aepipeline tool to reprocess and screen the data and the xselect tool to extract spectra, images and light curves. We used standard 
filtering, with the exception that we eliminated times with geomagnetic cut-off rigidity (COR2) less than 6 and time since the South Atlantic Anomaly 
(SAA) passage less than 436 sec. We excluded regions contaminated by the \element[][55]{Fe} calibration sources and circular regions with a 
radius of 1 arcmin centred on bright point sources seen in XMM-Newton images (see Fig. \ref{fig:xis1_reg}).

We constructed energy redistribution and auxiliary response files with the xisrmfgen and xissimarfgen tools respectively. 
For each cluster, we used the XMM-Newton / EPIC-pn count rate map to weight the position-dependent effective area with the 
brightness distribution when producing the auxiliary responses. For the sky background modelling we produced the auxiliary 
responses assuming a circular source of uniform brightness with a radius of 20 arcmin.
We used makepi files version 20070124 and 20100929, RMF parameter file 
version 20080901 and contamination file version 20091201 as included in XIS CALDB release 20110608. 
 
We checked XIS1 gain by searching for residuals in the spectral fits (Section \ref{sect:spect_analysis}) 
around the \ion{Fe}{xxv} lines. We found evidence for residuals in the Centaurus cluster and performed a gain fit in Xspec. However, the 
cluster temperature was unaffected.

\subsection{XMM-Newton EPIC-pn}

We used SAS version xmmsas\_20110223\_1801-11.0.0 and the latest calibration information in June 2011 for EPIC-pn data reduction. We used the epchain tool 
with the default parameters to produce event files. We filtered the event files, excluding bad pixels and CCD gaps (i.e. using expression flag=0), 
including only patterns 0 -- 4. We generated simulated out-of-time event files, which we used to subtract events registered 
during CCD readout according to pn mode (Table \ref{tab:obs_prop}).  

We used the evselect tool to extract spectra, images and light curves from the event file, excluding regions 
contaminated by bright point sources. We used the light curves to further filter the events by excluding the periods when the E $>$ 10 keV band flux 
exceeded the quiescent rate by $\sim$ 20\%
We used the rmfgen and arfgen tools to produce energy redistribution and auxiliary 
response files respectively. We used the arfgen tool in extended source configuration and weighted the response with an XMM-Newton
image of the cluster in detector coordinates, binned in 0.5 arcmin pixels.

\section{Background modelling}
\label{sect:bkg}

As our sample consists of nearby clusters that fill the whole FOV, local estimates for the background emission are 
not available.

\subsection{Sky background}
\label{skybkg}
We used the ROSAT All Sky Survey based HEASARC X-ray background tool 
\footnote{http://heasarc.gsfc.nasa.gov/cgi-bin/Tools/xraybg/xraybg.pl} 
to estimate the spectra of the sky background for both XMM-Newton / EPIC and Suzaku / XIS instruments. 
We extracted spectra of annular fields centred on the clusters, with an inner radius of 0.5 degrees and outer radius of 
1.0 degrees. We fitted the spectra with a model consisting of an absorbed and unabsorbed thermal MEKAL emission component 
with solar abundances \citep{mekal}, 
describing emission from the Galactic halo and Local Hot Bubble respectively, and an absorbed power-law with the 
index fixed to 1.45, describing the extragalactic X-ray background.\footnote{See e.g. http://heasarc.gsfc.nasa.gov/Tools/xraybg\_help.html for references.}
We fixed the absorption column densities to the value of the respective clusters (see Section \ref{sect:spect_analysis}). 
We added the local sky background emission model to all subsequent spectral fits. In order to 
account for the uncertainty in the sky background model, we allowed the temperatures and normalisations of the emission components to vary within 1.0$\sigma$ 
of their best-fit values during subsequent spectral fits.

\subsection{XMM-Newton non X-ray background}
\label{sect:XMM_nxb}
We extracted and investigated full FOV hard energy band (10 -- 14 keV) EPIC pn light curves to determine the 
quiescent non-X-ray background levels and performed GTI filtering by excluding flares. We modelled the quiescent particle 
background by scaling a sample of closed filter spectra \citep{Nevalainen05} to match hard band full FOV 
spectra. We removed the scaled closed filter spectra from the cluster spectra by using them as correction files is Xspec 
in subsequent pn spectral fits.

We studied the uncertainty of the adopted particle background model by varying the scaling of the closed filter spectra 
by $\pm$ 10 \% and interpreted the change in best fit temperature as a systematic effect due to background uncertainties.
This had on average a $\sim$ 3 \% effect on the best fit temperatures in the hard band; in the soft band, the effect was less than 
0.1 \% for all clusters in our sample. We propagated this uncertainty by adding it in quadrature to the statistical uncertainties 
of the temperatures.

\subsection{Suzaku non X-ray background}
\label{sect:suzaku_nxb}
We used the xisnxbgen tool with default settings to generate spectra of the Suzaku XIS non X-ray background (NXB) based 
on night Earth observations. We used COR2 weighting for NXB construction and to verify the normalisation of the NXB 
generated using the xisnxbgen tool, we compared the $>$ 10 keV NXB flux with that in the cluster data and found them 
consistent. We subsequently subtracted the NXB spectra from the cluster spectra 
in all Suzaku XIS spectral fits by using the NXB spectra as background files in Xspec and thus 
considered the uncertainty of the NXB in the spectral fits. 

\begin{figure}
  \resizebox{\hsize}{!}{\includegraphics{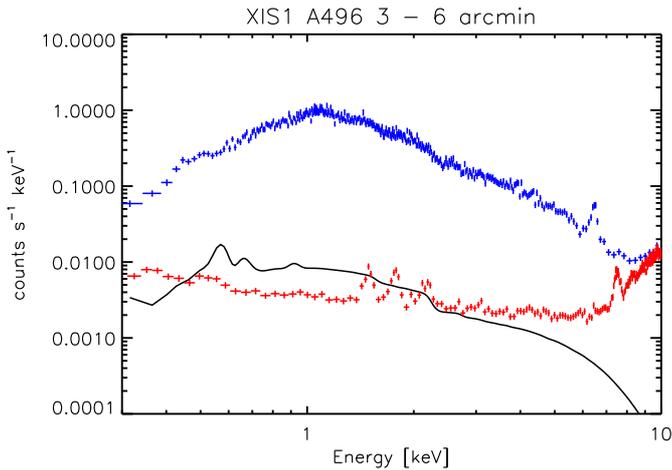}}
\caption{The total emission (blue crosses), sky background (black curve) and non X-ray background (red crosses) emission 
of A496 in the 3 -- 6 arcmin region for XIS1. The sky background is modelled from ROSAT All Sky Survey data and folded 
through the XIS1 response. The XIS1 observation of A496 has the strongest relative background emission 
of the cluster sample.
}
\label{fig:bkg}
\end{figure}

\subsection{Charge exchange contamination}
\label{sect:trans_app}
Geocoronal and heliospheric Solar wind charge exchange (SWCX) \citep{wargelin04, snowden04, fujimoto07} may produce 
contamination to the cluster emission in terms of emission lines in the 0.5 -- 0.9 keV band. The total flux of the SWCX 
varies by an order of magnitude at a timescale of hours or minutes, reaching a maximum level comparable to that of the 
sky background (Galactic emission + CXB). To obtain an conservative estimate of the possible SWCX
effect on our results, we examined the variation of the best-fit 0.5 -- 2.0 keV energy band temperatures of XIS1 when 
keeping the sky background model fixed to RASS values, and by increasing the normalisation by 100 \%. Due to the low 
level of the sky background compared to the cluster flux in our sample the above effect was insignificant.

\subsection{XIS point spread function}
\label{sect:PSF}

Cluster flux originating from a given extraction region may be contaminated by PSF scattered flux from neighbouring 
regions \citep[e.g.][]{Sato07}. Since Suzaku XIS has a relatively large PSF ($\sim$ 2 arcmin half-power diameter), scatter from cool 
cores may affect observed XIS temperatures. Thus we examined how well PSF scatter to and from our extraction region (3 -- 6 arcmin 
annulus, see Section \ref{sect:regions}) is minimised by the usage of extraction regions wider and further away from the 
centre than the XIS PSF.

The exact PSF scattering effects depend on the brightness of the cool centre, the shape of the surface density profile, 
the temperature of the cluster and the cluster redshift, and therefore they can vary from cluster to cluster. We choose to use A1795 
and A3112 as conservative examples for the PSF-induced systematic uncertainty, as they 
possess cool cores which are an order of magnitude brighter than the cores of the rest of the sample \citep{Peres98}.

We performed ray-tracing simulations of A1795 with xissim 
to estimate the fraction of photons that originate from the central cool region (r $\sim$ 1.5 arcmin; \citet{Vikhlinin05}) but are 
scattered into the 3 -- 6 arcmin annulus.
We used an exposure corrected 0.5 -- 7.5 keV energy band XMM-Newton EPIC / pn rate map of A1795 as a source 
brightness model and, since the energy dependence of the PSF is known to be small (e.g. \citet{Sato07} Appendix 1), 
performed the simulation at 1 keV. We accounted for typical pointing errors by using the attitude and
exposure information from the Suzaku observations of A1795 used in this work.  The simulations show that about 16 \% 
of the flux detected in the 3 -- 6 arcmin extraction region originates from the inner 1.5 arcmin cool core.
There is a small variation of scattered light between detectors, with values of 13 \% (XIS0), 18 \% (XIS1), and 
15 \% (XIS0).

We then simulated a XIS1 spectrum in the 3 -- 6 arcmin annulus containing one MEKAL component for the emission originating 
from that region and another for the PSF scattered emission from the cool core. We used the Chandra measurements of A1795 
\citep{Vikhlinin05} for the temperature and abundance of the two components:  kT = 6.5 keV, 0.25 solar and kT = 4.75 keV,  
0.55 solar, respectively. We used relative model normalisations of 0.84 and 0.16 respectively, given by the above PSF 
simulations. We fitted the simulated spectrum with a single temperature MEKAL model in the soft 0.5 -- 2.0 keV and hard 
2.0 -- 7.0 keV energy bands. The resulting best-fit temperatures were lower by $\sim$ 7 \% and $\sim$ 6 \% in the soft 
and hard energy bands respectively, compared to those obtained from the fits to the observational data.

\citet{Lehto10} performed an analogous 
XIS PSF scatter ray-tracing simulation on A3112 data using a similar 3 -- 6 arcmin extraction region as used for our study. 
They found that $\sim$ 10 \% of the flux in the extraction region originates from the core, but best-fit temperatures were 
consistent within 2 \% in all energy bands studied. The other clusters are weaker (or non-) cool 
cores, so we therefore expect lower biases for all of them. Thus we conclude that cool core contributions may bias Suzaku XIS
temperatures low, but only up to 6 --7 \% in the worst case, and much lower in most cases. 
However, as the three XIS instruments have similar scattered light
properties, they would be biased by similar amounts and the relative XIS 
temperature differences would be unaffected. Consequently, the effect is only limited to 
XIS to pn temperature comparisons.

\section{Methods}
\label{sect:spect_analysis}
\citet{nevalainen10} found that several independent missions / instruments  (XMM-Newton / EPIC, Chandra / ACIS, BeppoSAX / MECS)
and two independent physical processes (bremsstrahlung and collisionally excited line emission) yielded consistent temperatures  in the 2 -- 7 keV band.
However, \citet{nevalainen10} also found that the 2 -- 7 keV band EPIC / ACIS fluxes disagree by 5 -- 10 \%. Here we test the above findings with
Suzaku / XIS by studying the cross-calibration of the the effective area by spectroscopic analysis of cluster data obtained 
with Suzaku / XIS instruments and the XMM-Newton / EPIC-pn instrument.

\subsection{Spectral analysis}

We extracted spectra of the clusters in our sample from annular 3 -- 6 arcmin regions centred on the clusters 
(see Section \ref{sect:regions}). We binned the spectra to a minimum of 100 counts per bin. 
We fitted the NXB-subtracted spectra (see Sections \ref{sect:suzaku_nxb} and \ref{sect:XMM_nxb}) with a model consisting of an absorbed single temperature 
MEKAL model \citep{mekal} and the sky background component (see Section \ref{skybkg}).
We used metal abundances of \citet{grsa}. We modelled the Galactic absorption with the PHABS model, fixing the column densities to the values 
obtained with 21 cm radio observations \citep{kalberla05} and using the absorption cross sections of \citet{b-c92}.
We allowed the temperature, metal abundance and normalisation of the cluster emission model to vary in the fits. We performed the fits in two 
energy bands: 0.5 -- 2.0 keV (soft band) and 2.0 -- 7.0 keV (hard band). 

In addition to excluding clusters with N$_{\rm H}$  $> 6 \times 10^{20}$ cm$^2$ from the soft band sample (Section \ref{sect:sample}), we excluded Coma from the 
soft band sample as we were not able to produce single temperature spectral fits of satisfactory statistical quality. Thus the soft band sample contains 
a smaller number of objects than the hard band sample. The properties of the observations are presented in Table \ref{tab:obs}.

\begin{table*}
\caption{Properties of the observations} 
\label{tab:obs} 
\centering 
\begin{tabular}{l|c c c c|c c c c} 
\hline\hline 
Name & \multicolumn{4}{c|}{Background\tablefootmark{a} / total flux} & \multicolumn{4}{c}{Total data counts} \\
     & pn   & XIS0 & XIS1 & XIS3 & pn    & XIS0   & XIS1   & XIS3   \\
\hline
\multicolumn{1}{c}{} & \multicolumn{8}{c}{Soft energy band (0.5 -- 2.0 keV)} \\
\hline
A1060       & 0.04 & 0.03 & 0.03 & 0.03 & 73473 & 24723 & 38605 & 22610 \\
A1795       & 0.04 & 0.05 & 0.06 & 0.06 & 149804& 8296  & 13926 & 7537  \\
A262        & 0.05 & 0.04 & 0.05 & 0.04 & 40870 & 14479 & 21265 & 13668 \\
A3112       & 0.08 & 0.07 & 0.07 & 0.08 & 95084 & 10865 & 17928 & 11140 \\
A496        & 0.02 & 0.08 & 0.10 & 0.09 & 134661& 20235 & 29516 & 18686 \\
\hline 
\multicolumn{1}{c}{} & \multicolumn{8}{c}{Hard energy band (2.0 -- 7.0 keV)} \\
\hline
A1060       & 0.04 & 0.03 & 0.04 & 0.03 & 18789 & 12033 & 13982 & 11898 \\
A1795       & 0.02 & 0.04 & 0.04 & 0.04 & 48455 & 5034  & 6125  & 5148  \\
A262        & 0.07 & 0.06 & 0.06 & 0.07 & 8151  & 5958  & 6829  & 6829  \\
A3112       & 0.08 & 0.06 & 0.08 & 0.06 & 28344 & 7014  & 8628  & 7298  \\
A496        & 0.02 & 0.06 & 0.10 & 0.06 & 42088 & 13524 & 14770 & 12786 \\
AWM7        & 0.03 & 0.02 & 0.02 & 0.02 & 176775& 12786 & 12235 & 11008 \\
Centaurus   & 0.05 & 0.04 & 0.06 & 0.05 & 115211& 18242 & 20907 & 17953 \\
Coma        & 0.02 & 0.02 & 0.02 & 0.02 & 74184 & 220176& 242263& 229227\\
Ophiuchus   & 0.01 & 0.01 & 0.01 & 0.01 & 114581& 265097& 287684& 271563\\
Triangulum  & 0.02 & 0.01 & 0.01 & 0.02 & 11799 & 70564 & 74603 & 68502 \\
\hline 
\end{tabular}
\tablefoot{\tablefoottext{a}{total sky and non X-ray background}
}
\end{table*}

We experimented with using C statistics \citep{Cstat} in addition to standard $\chi^2$ statistics to estimate the best-fit parameters. 
We compared the temperatures obtained by fitting unbinned XIS1 spectra using C statistics to XIS1 temperatures obtained using $\chi^{2}$ statistics 
and binning the spectra to a minimum of 100 counts per bin. Using the statistical analysis described below, we found that the 
C statistics yields systematically higher temperatures, with an average temperature difference of 0.5 \% 
in the soft 0.5 -- 2.0 keV energy band (statistical significance 0.2$\sigma$) and 0.8 \% in the hard 2.0 -- 7.0 keV energy band 
(0.8$\sigma$). Thus the choice of statistics in spectral fits can have an impact on cluster temperatures, but it has a low 
statistical significance and the effect is consequently minimal due to the high number of counts in our spectra. We will use binned spectra and $\chi^{2}$ 
statistics for the remainder of this paper.

\subsection{Comparison of temperatures and fluxes}

We computed the difference in temperature or flux measurements and their statistical significance for each cluster and instrument 
pairing in terms of a fraction of the average value, $f_T = \Delta T/<T> = (T_1 - T_2) / \{(T_1 + T_2) / 2\}$ and $\sigma_T = \sigma_{\Delta T} /<T>$; 
$f_F = \Delta F/<F> = (F_1 - F_2) / \{(F_1 + F_2) / 2\}$ and $\sigma_F = \sigma_{\Delta F} /<F>$ respectively. 
As a systematic uncertainty in calibration would tend to drive the mean of the temperature or flux differences away from zero, 
we attributed $f_T$ or $f_F$ as a measure of the relative systematic uncertainty in the cross-calibration of a given instrument pair.

\subsection{Stacked residuals}
\label{sect:resid}

To examine the cross-calibration accuracy of the effective area of the studied instruments, we applied the stacked residuals method 
\citep{longinotti08}. This method involves a choice of a reference instrument, against which the other instruments are compared.
Since we do not know which, if any, instrument has its effective area perfectly calibrated,  the choice of a reference 
instrument is arbitrary and the comparison is only relative.

The stacked residuals method for a pair of instruments, I1 and I2, involves the following steps:

\begin{itemize}

\item
We start with the data (data$_{\rm ~I1, I2}$) and responses (resp$_{\rm ~I1, I2}$), defined here as the energy response matrix (RMF) multiplied 
with the effective area (ARF), of the instruments studied. For this example we use I1 as the reference
instrument and thus perform spectral analysis as explained above for each cluster to fit a reference model (model$_{\rm ~I1}$). 

\item
In order to be able to compare the data of different instruments we regroup data$_{\rm ~I1, I2}$ to equal binning

\item
For each source, we convolve the reference model with the response of the instruments in order to produce an I1-based model prediction  
(pred$_{\rm ~I1, I2}$ = model$_{\rm ~I1}$ $\otimes$ resp$_{\rm ~I1, I2}$)

\item 
We divide the data of each cluster by the above model prediction to obtain residuals over the I1-based prediction
(resid$_{\rm ~I1, I2}$ = data$_{\rm ~I1, I2}$ / pred$_{\rm ~I1, I2}$)

\item 
We divide the above residuals of I2 with I1 residuals, in order to remove effects of possible calibration uncertainties of I1,
obtaining ratio R$_{21}$ for I2:
\begin{eqnarray}
{\rm R}_{21} &  = & {\rm resid}_{\rm ~I2}~ / ~{\rm resid}_{\rm ~I1} \nonumber \\ 
 &  = &  \frac{\rm data_{\rm ~I2}}{\rm model_{\rm ~I1} \otimes {\rm resp}_{\rm ~I2}} \times \frac{\rm model_{\rm ~I1} \otimes {\rm resp}_{\rm ~I1}}{\rm data_{\rm ~I1} }
\label{R0131}
\end{eqnarray}
R$_{21}$ is a useful measure of the effective area cross-calibration, since the value for each energy bin would be unity if the cross-calibration of I1 and I2 effective areas is consistent.

\item
We calculate median and median absolute deviation of R$_{21}$  values for each energy channel for the cluster sample.

\end{itemize}

\section{Results}

The best-fit single temperature models, data and spectral parameters are shown in Appendix \ref{sect:spect_fits}.

\subsection{Hard band} 
\label{sect:new_xis_hb}

\begin{figure*}
\centering
\includegraphics[width=17cm]{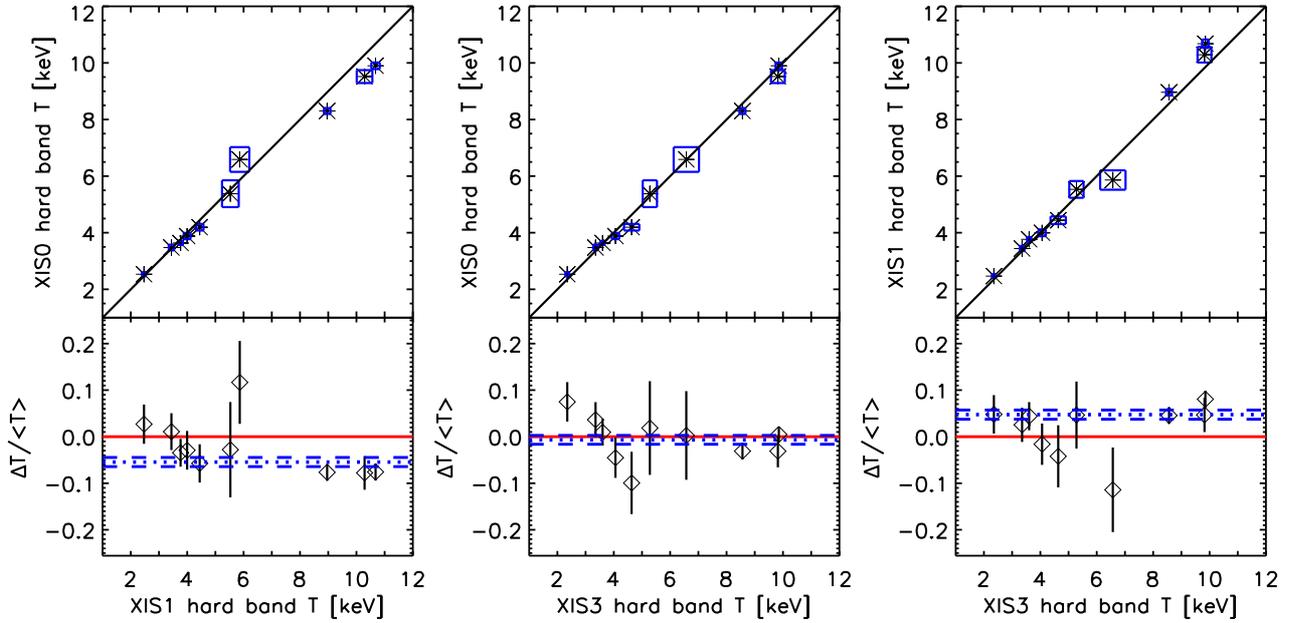}
\vspace{-3.5cm}
\caption{
{\it Upper panel:} The best-fit hard band temperatures (asterisks) and 1$\sigma$ uncertainties (boxes) for Suzaku XIS instruments using the public calibration. The solid black line is an identity line, drawn as a reference.
{\it Lower panel:} The relative temperature differences $f_{T}$ (diamonds) between different Suzaku XIS instruments, and their 1$\sigma$ uncertainties. The dotted and dashed lines show the weighted means of the relative temperature difference and corresponding uncertainties.
\label{fig:new_xis_hb}}
\end{figure*}

We found that the hard band (2.0 -- 7.0 keV) temperatures measured with XIS0 and XIS3 agree, with an average difference of $\sim$ 1 \% 
(statistical significance 0.7$\sigma$). Temperatures 
measured with XIS1 are on average $\sim$ 5 \% (5.6$\sigma$) and $\sim$ 5 \% (4.9$\sigma$) higher than those measured 
with XIS0 and XIS3 respectively (see Fig. \ref{fig:new_xis_hb}). 

The simultaneous Suzaku / XMM-Newton observations of BL Lac object PKS2155-304 \citep{Ishida11} are useful for comparison with our 
cluster results since the observations cover the same time period (2005 -- 2008).  The XIS CALDB version used by \citet{Ishida11} for PKS2155-304 
(dated July 30, 2010) differs from the one used for our cluster study (dated June 8, 2011) in the model used for the molecular OBF contamination 
(see Section 8).  Thus there should be little difference in the calibration in the hard band.  The best-fit flux ratios of PKS2155-304 indicate that  
XIS1 yields somewhat harder spectra in the 2 -- 7 keV band than XIS0, consistent with our results.

\begin{figure}
  \resizebox{\hsize}{!}{\includegraphics{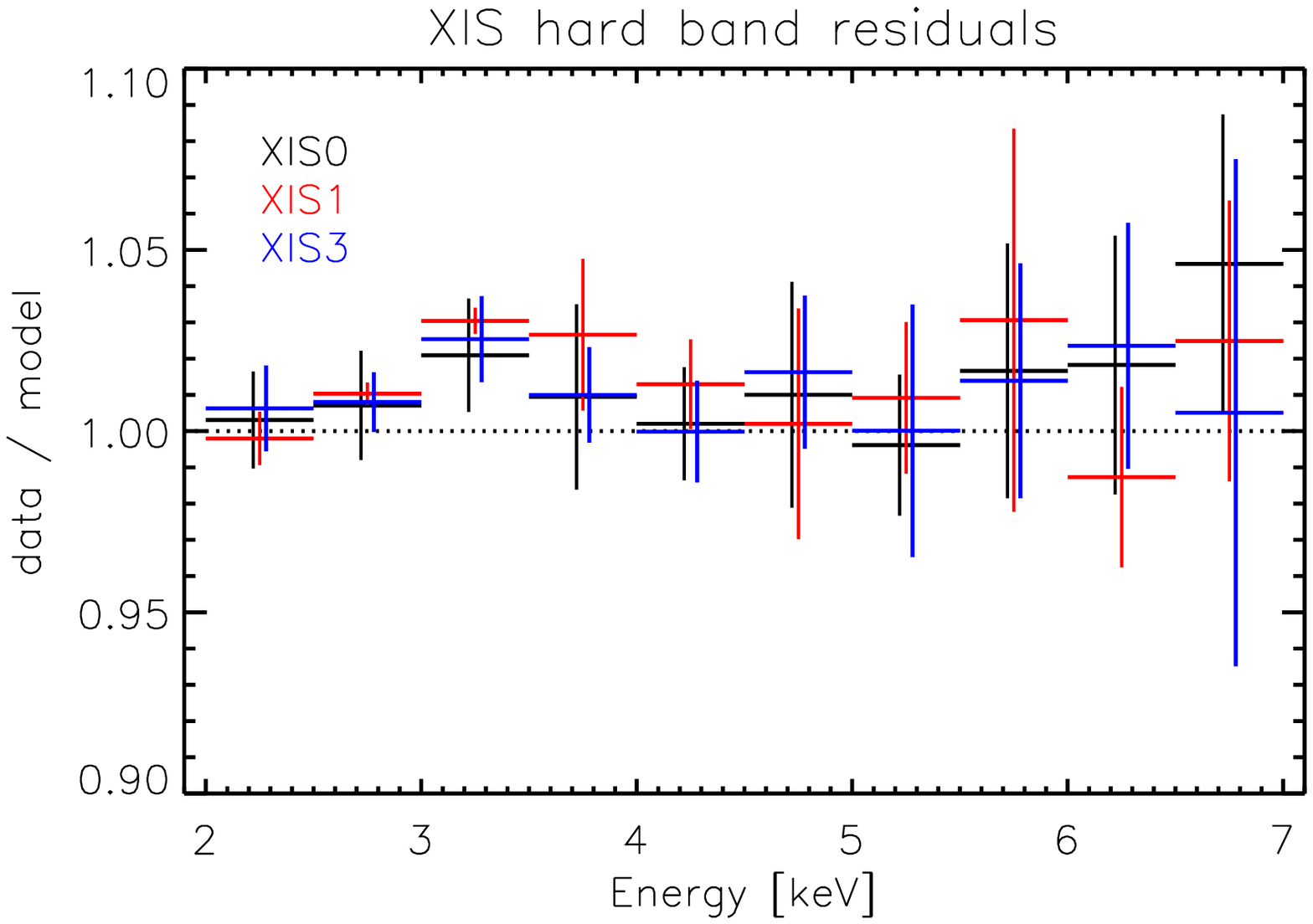}}
\caption{The median and the median absolute deviation of the hard band residuals (data / best-fit model) for XIS0 (top), XIS1 (middle) and XIS3 (bottom). The model is an absorbed single temperature MEKAL whose temperatures and metal abundances are independent for each instrument, fitted to 2.0 -- 7.0 keV band using the public calibration. The data are regrouped to uniform binning of 0.5 keV.
}
\label{fig:xis_hb_res}
\end{figure}

The hard band best-fit residuals of the clusters using the XIS instruments are rather consistent with unity (see Fig. \ref{fig:xis_hb_res}) except that there are some systematic residuals around  3.0 -- 3.5 keV energies at a few \% level, especially in the XIS1 data.
These features are probably causing the small differences in XIS hard band temperatures (see above). This behaviour is possibly connected to the PSF effect, see below.

\begin{figure*}
\centering
\includegraphics[width=17cm]{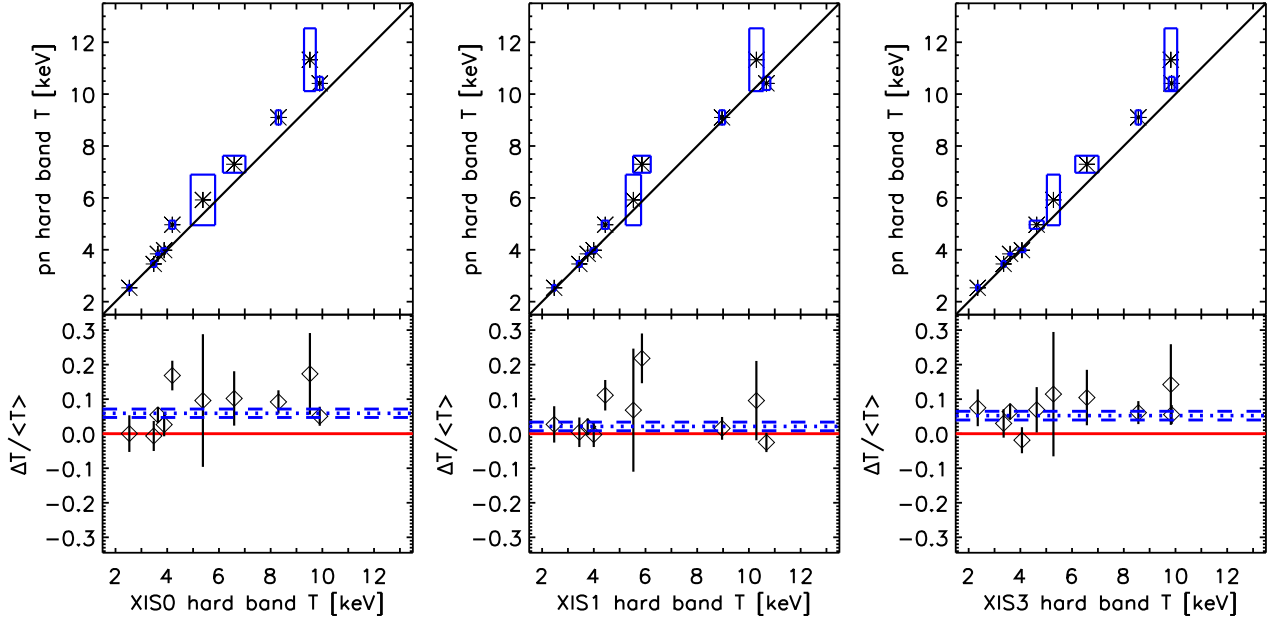}
\vspace{-3.5cm}
\caption{
{\it Upper panel:} The best-fit hard band temperatures (asterisks) and 1$\sigma$ uncertainties (boxes) for XMM-Newton EPIC-pn and Suzaku XIS instruments using the public calibration. The solid black line is an identity line, drawn as a reference.
{\it Lower panel:} The relative temperature differences $f_{T}$ (diamonds) between different XMM-Newton EPIC-pn and Suzaku XIS instruments, and their 1$\sigma$ uncertainties. The dotted and dashed lines show the weighted means of the relative temperature difference and corresponding uncertainties.
\label{fig:pn_xis_new_hb}}
\end{figure*}

\begin{figure}
  \resizebox{\hsize}{!}{\includegraphics{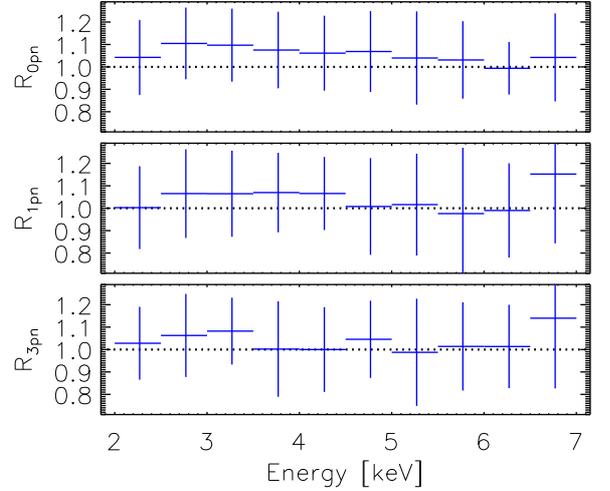}}
    \caption{The median $\pm$ median absolute deviation of hard energy band stacked residuals of XIS0 / pn R$_{\rm 0pn}$ (top panel), XIS1 / pn 
R$_{\rm 1pn}$ (middle panel) and XIS3 / pn R$_{\rm 3pn}$ (bottom panel). The data are regrouped to uniform binning of 0.5 keV.
}
    \label{fig:xis_pn_2070_rat}
\end{figure}

In order to derive the effects of possible cross-calibration uncertainties in the shape of the effective area in the hard band between Suzaku-XIS and EPIC-pn instruments,   
 we compared the XIS hard band temperatures of our sample to those obtained with EPIC-pn. 
We found that the XIS1 and EPIC-pn temperatures are consistent within the statistical uncertainties: on average the XIS1 temperatures are $\sim$ 2 \%  
(1.7$\sigma$) lower than the EPIC-pn values. XIS0 and XIS3 temperatures are respectively on average $\sim$ 6 \% (4.9$\sigma$) and  $\sim$ 5 \% 
(4.2$\sigma$) lower than pn temperatures (see Fig. \ref{fig:pn_xis_new_hb}), indicating some cross-calibration problems.
However, the lower XIS temperatures may be due to the uncorrected PSF scattering from the cool core which may bias XIS temperatures low by a maximum of $\sim$ 6 \% in the hard band (see Section \ref{sect:PSF}). 
Indeed, the stacked XIS / pn residuals indicate some excess data at the lowest energies, reaching a maximum of $\sim$10\% at $\sim$ 3 keV energies, which is possibly due to the PSF scatter from the cool core (see Fig. \ref{fig:xis_pn_2070_rat}). 
Since the effective area energy dependence of XMM-Newton's EPIC-pn is most likely well calibrated in the hard band \citep{nevalainen10}, 
the shape of the hard band effective areas of the XIS instruments are correctly calibrated within the statistical and PSF-related uncertainties, amounting to $\sim$ 5\% uncertainties on the measurement 
of the hard band temperature.

\subsection{Soft band}
\label{sect:new_xis_sb}

The temperature measurement of hot clusters is not optimally achieved in the 0.5 -- 2.0 keV band as the temperature dependence of the continuum 
shape is relatively weak since the cut-off is at higher energies. Due to weak line emission in the soft energy band at typical cluster temperatures, 
possible uncertainties in metal abundance measurements mostly produce variation in continuum emission, which in turn may affect the temperature measurements. 
However, our aim here is to yield a phenomenological, rather than physical model of the combination of cluster 
continuum emission and effective area calibration uncertainties. Thus the soft band temperatures are still useful for the purposes of cross-calibration 
of the effective area.

\subsubsection{\textbf{Temperatures and residuals}}
\label{public_t}
\begin{figure*}
\centering
\includegraphics[width=17cm]{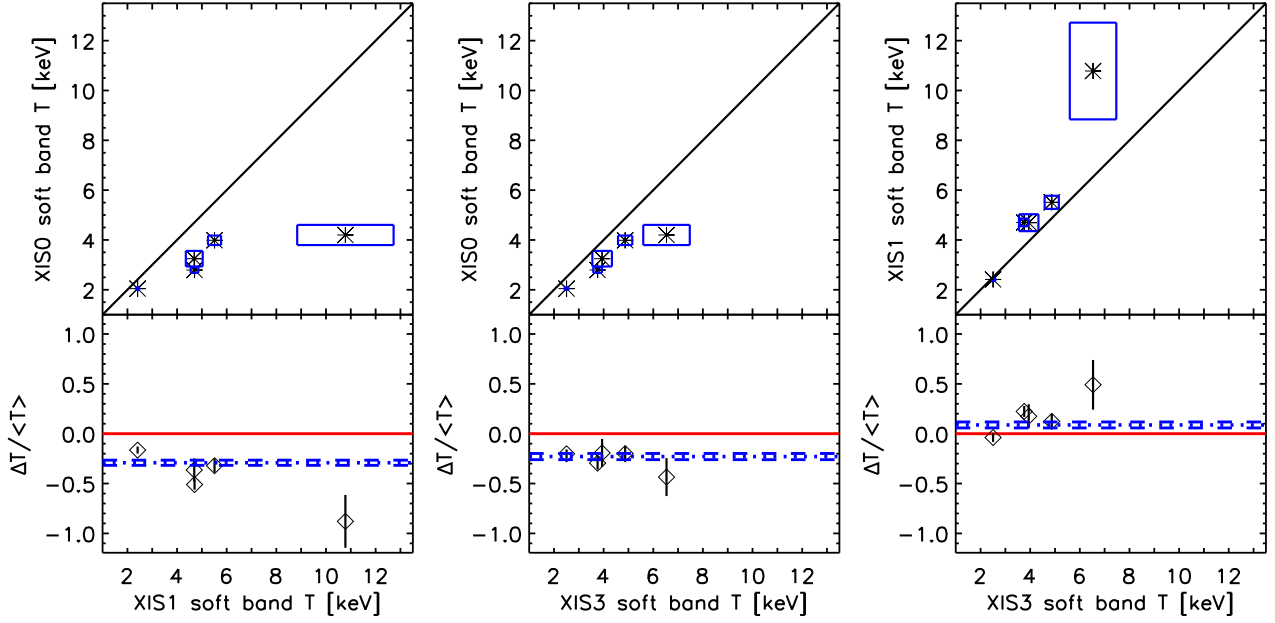}
\vspace{-3.5cm}
\caption{
{\it Upper panel:} The best-fit soft band temperatures (asterisks) and 1$\sigma$ uncertainties (boxes) for Suzaku XIS instruments using the public
calibration. The solid black line is an identity line, drawn as a reference.
{\it Lower panel:} The relative temperature differences $f_{T}$ (diamonds) between different Suzaku XIS instruments, and their 1$\sigma$ uncertainties. The dotted and dashed lines show the weighted means of the relative temperature difference and corresponding uncertainties.
}
\label{fig:new_xis_sb}
\end{figure*}

We fitted the soft band spectra of the clusters in the soft band sample with an absorbed single-temperature MEKAL model (see Table \ref{tab:fit_par}). 
The soft band temperatures obtained using XIS1 and XIS3 differ by $\sim$ 9 \% , which is not very significant 
(2.9$\sigma$). Unfortunately XIS0 temperatures disagree significantly from the other two instruments. They are on average 
$\sim$ 29 \% (11.2$\sigma$) and $\sim$ 23 \% (7.4$\sigma$) lower than XIS1 and XIS3 temperatures respectively 
(see Fig. \ref{fig:new_xis_sb}). 

The \citet{Ishida11} work on PKS2155-304 also indicates that XIS0 yields softer spectra than XIS1 in the 0.5 -- 2.0 keV band during years 2005 -- 2008, qualitatively consistent with our cluster results.  However, as noted in Section 7.1, 
\citet{Ishida11} used an older XIS calibration with a different model for the OBF contaminant, making a more rigorous 
comparison to that work difficult.

\begin{figure}
  \resizebox{\hsize}{!}{\includegraphics{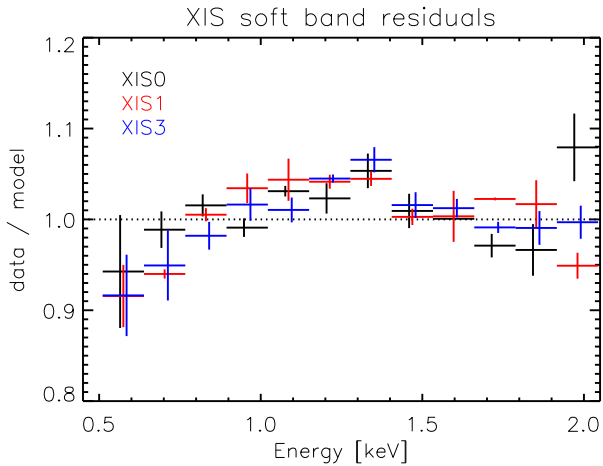}}
    \caption{The median and the median absolute deviation of the residuals (data / best-fit model) of the soft band cluster sample for XIS0 (black), 
XIS1 (red) and XIS3 (blue). The model is an absorbed single temperature MEKAL whose temperatures and metal abundances are independent for each instrument, 
fitted to 0.5 -- 2.0 keV band using the public calibration. The data are regrouped to uniform binning of 127.75 keV.
}
    \label{fig:xis_sb_residuals}
\end{figure}

XIS soft band residuals exhibit a systematic behaviour: they reach a maximum at 1.3 -- 1.4 
keV while they decrease with lower and higher energy (see Fig. \ref{fig:xis_sb_residuals}). We attempted two-temperature modelling, but this did not 
significantly improve the spectral fits. Importantly, the residuals of the XIS0 unit, which yielded significantly lower soft 
band temperatures (see above) are rather consistent with unity below 1 keV energies, while those of XIS1 and XIS3 decrease significantly 
below unity. Assuming for a moment that the shape of the XIS0 effective area below 1 keV is accurately calibrated, this indicates that the effective areas of XIS1 and XIS3 are overestimated increasingly with lower energy.

\begin{figure*}
\centering
\includegraphics[width=17cm]{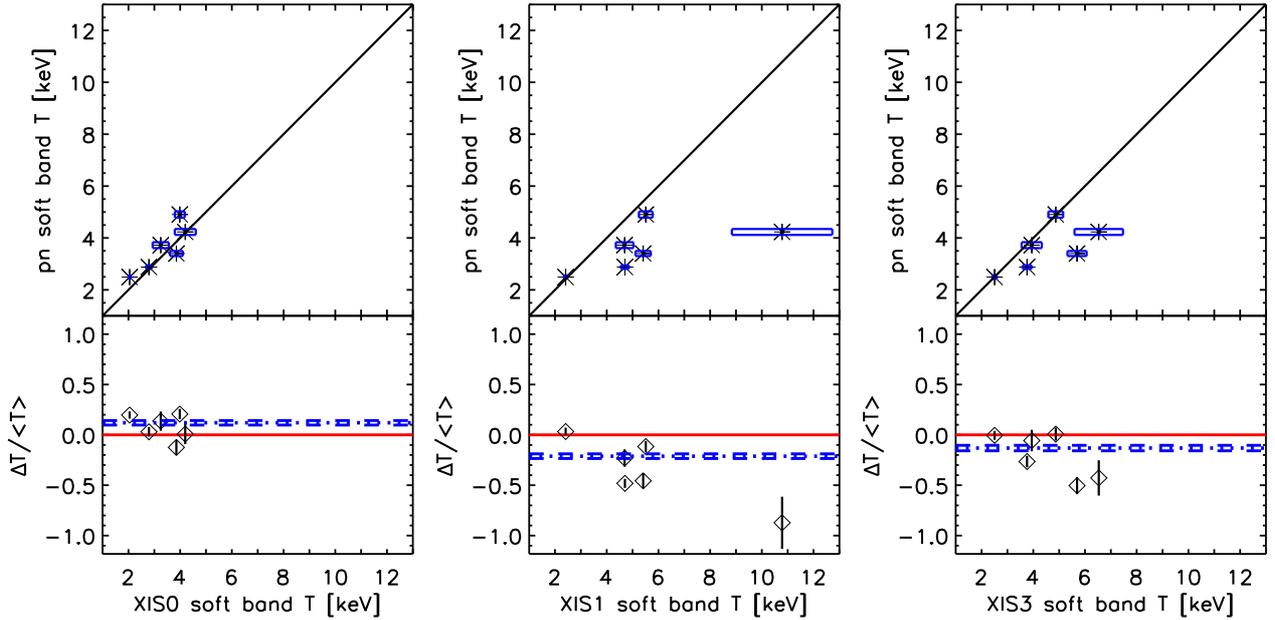}
\vspace{-3.5cm}
\caption{
{\it Upper panel:} The best-fit soft band temperatures (asterisks) and 1$\sigma$ uncertainties (boxes) for  XMM-Newton EPIC-pn and Suzaku XIS instruments using the public calibration. The solid black line is an identity line, drawn as a reference.
{\it Lower panel:} The relative temperature differences $f_{T}$ (diamonds) between XMM-Newton EPIC-pn and Suzaku XIS instruments, and their 1$\sigma$ uncertainties. The dotted and dashed lines show the weighted means of the relative temperature difference and corresponding uncertainties.
\label{fig:pn_xis_new_sb}}
\end{figure*}

The comparison between XIS and XMM-Newton / pn soft band temperatures complicates the situation further:
XIS0 values are on average $\sim$ 14 \% (6.2$\sigma$) lower than those of pn, whereas XIS1 and XIS3 values are respectively on 
average $\sim$ 18 \% (7.4$\sigma$) and $\sim$ 9 \% (3.0$\sigma$) higher than those of pn (see Fig. \ref{fig:pn_xis_new_sb}). 
Since PSF scatter from the cool core could bias XIS temperatures low by 7 \% in only the worst case, this effect is unable to fully account
for the XIS0 / pn temperature difference, and in fact any correction for it would exacerbate XIS1 / pn and XIS3 / pn differences.

The observations of PKS2155-304 \citep{Ishida11} also indicate that XIS1 yields harder spectra than pn in the 0.5 -- 2.0 
keV energy band. As Chandra / ACIS soft band cluster temperatures are on average $\sim$ 20 \% higher than pn soft band 
temperatures \citep{nevalainen10}, XIS1 / ACIS are the only instrument pair that yields approximatively 
consistent sample average soft band temperatures. Due to these disagreements we do not have clear indications of which 
instruments are most accurately calibrated in the soft energy band.

\subsubsection{Stacked residuals}
\label{effarea}

\begin{figure*}
\centering
\includegraphics[width=17cm]{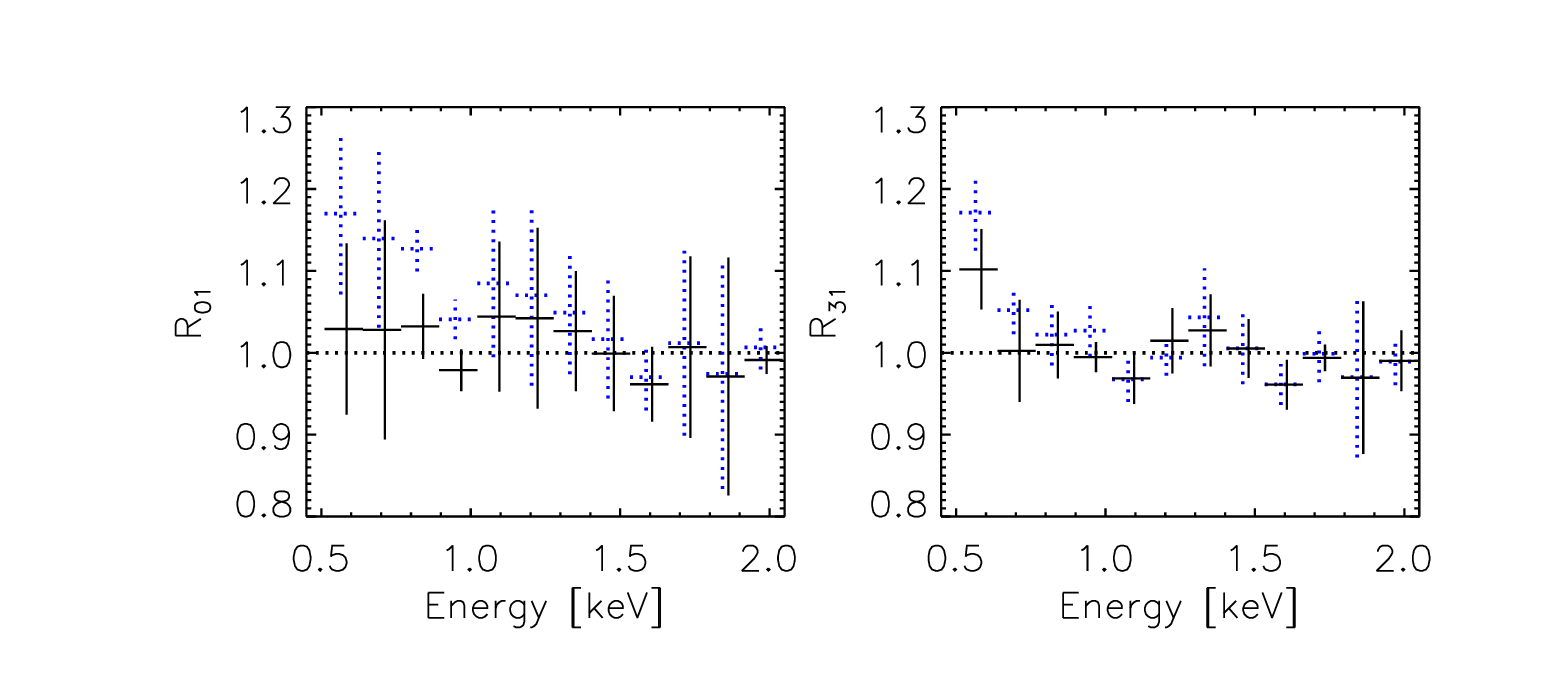}
\caption{The median $\pm$ median absolute deviation of soft energy band stacked residuals of XIS0 / XIS1 R$_{01}$ (left panel) and XIS3 / XIS1 R$_{31}$ 
(right panel) using the public calibration (dotted line) and the modified responses described in this work (solid line). 
The data are regrouped to uniform binning of 127.75 keV.
}
\label{fig:xis_mekalmod_rat_simult0520_sys_med}
\end{figure*}

However, this unfortunate situation does not prohibit us from deriving information about XIS cross-calibration by analysing the cluster 
measurements. The above XIS temperature and residual discrepancies are indicative of uncertainties of the cross-calibration of the energy dependence of the 
effective area of different XIS instruments. 
As XIS1 as a back-illuminated (BI) device has the largest effective area of the three XIS CCDs, we adopted it as the reference instrument 
for a relative cross-calibration comparison between the XIS instruments using the stacked residuals method (see Section \ref{sect:resid}).
The analysis indicated a systematic difference between the cross-calibration of XIS0 / XIS1 effective areas, which increases towards lower energies (see  Fig. \ref{fig:xis_mekalmod_rat_simult0520_sys_med}). The median of XIS0 / XIS1 residuals reaches a value of $\sim$ 1.2 at 0.5 keV, indicating 
$\sim$ 20 \% cross-calibration uncertainties. The case of XIS3 / XIS1 is less problematic, since the median of XIS3 / XIS1 differs significantly from 
unity only in the energy bin at $\sim$ 0.5 -- 0.7 keV. 

\begin{figure}
  \resizebox{\hsize}{!}{\includegraphics{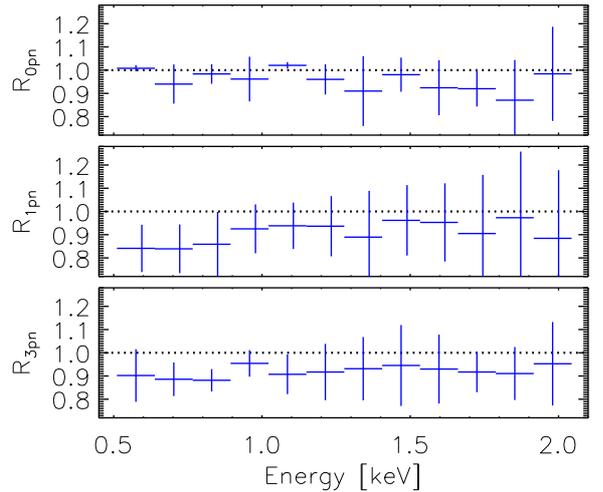}}
    \caption{The median $\pm$ median absolute deviation of the soft energy band stacked residuals of XIS0 / pn R$_{\rm 0pn}$ (top panel), XIS1 / pn 
R$_{\rm 1pn}$ (middle panel) and XIS3 / pn R$_{\rm 3pn}$ (bottom panel). The data are regrouped to uniform binning of 127.75 keV.}
    \label{fig:pn_xis_mekalmod_rat_2070_sys_med}
\end{figure}

In order to evaluate the cross-calibration uncertainties of the soft band effective area shape between Suzaku / XIS
 and XMM-Newton / pn instruments, indicated by the differences of the soft band temperatures (see Section 
\ref{public_t}), we calculated the stacked residuals using EPIC-pn as the reference instrument 
(Fig. \ref{fig:pn_xis_mekalmod_rat_2070_sys_med}). While XIS0 / pn residuals are rather consistent with unity at the lowest 
energies, the data indicates that at energies above 1.3 keV  XIS0 flux is lower than EPIC-pn flux by $\sim$ 10 \%. 
This effect would account for the systematically lower soft band XIS0 temperatures, when compared to EPIC-pn. 
On the other hand, XIS1 and XIS3 instruments 
yield $\sim$ 10 -- 20 \% lower flux than EPIC-pn at energies below 1 keV. This results in higher soft band temperatures 
for XIS1 and XIS3, when compared to EPIC-pn.

However, we do not have indications of which instrument is most accurately calibrated in the soft band. We thus 
conclude that if fitting only soft band data (0.5--2.0 keV), as is often done for high redshift clusters whose emission 
in the hard band is small compared to the background, using Suzaku-XIS or EPIC-pn instruments, the cross-calibration 
uncertainties between these instruments yield a maximum systematic uncertainty of $\sim$ 30\% on the derived temperatures.

\subsection{Full band}
\label{sect:full_band}
\begin{figure}
  \resizebox{\hsize}{!}{\includegraphics{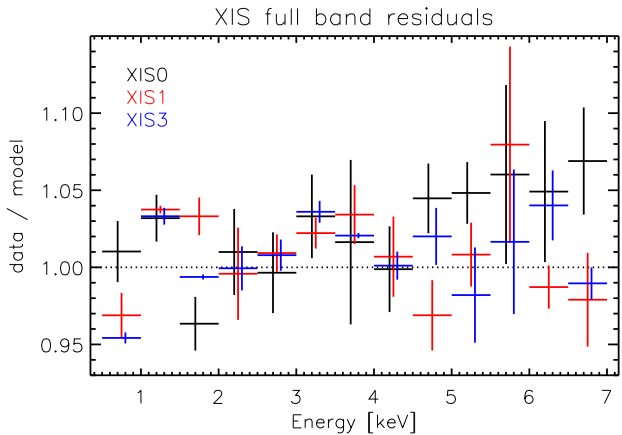}}
    \caption{The median and the median absolute deviation of the residuals (data / best-fit model) for the soft band cluster sample when 
fitting the full 0.5--7.0 keV band using the public calibration for XIS0 (black), 
XIS1 (red) and XIS3 (blue). The model is an absorbed single temperature MEKAL whose temperatures and metal abundances are independent for 
each instrument. The data are regrouped to uniform binning of 0.5 keV.
} 
   \label{fig:xis_fb_residuals}
\end{figure}

We then proceeded to evaluate the effects of the above cross-calibration uncertainties on standard astrophysical 
spectral analysis of the cluster spectra, i.e. modelling and analysis of the full useful energy band of 0.5 -- 7.0 keV,
using the soft band sample. These results should be applicable for any type of source which produces such 
continuum-dominated X-ray emission in this energy band which has significant statistical weight at energies below 2 keV.
However, the usage of a singe-temperature emission model in the full band may not be physically justified due to 
possible multi-temperature structure of the intracluster medium. Indeed, while the XIS residuals show systematic features 
due to the calibration uncertainties discussed above, a part of the residuals may originate from  multi-temperature 
structures (see Fig. \ref{fig:xis_fb_residuals}).
Yet the usage of the single-temperature emission model has the advantage of yielding a simple and useful measure for 
cross-calibration uncertainties.

\subsubsection{Temperatures}

While the cross-calibration uncertainties between XIS instruments and between Suzaku-XIS and EPIC-pn instruments were quite large below 2 keV (see Section \ref{sect:new_xis_sb}), the largest discrepancies were concentrated in a rather narrow energy band of 0.5 -- 0.8 keV (see Figs. \ref{fig:xis_mekalmod_rat_simult0520_sys_med}, \ref{fig:pn_xis_mekalmod_rat_2070_sys_med}). The emission of this band is relatively small when compared to the total emission of the full band 
and thus the effect of the soft band discrepancies on full band temperatures is small. The full band analysis 
resulted in rather similar temperatures as the well calibrated hard band.\footnote{The differences in the hard and wide band samples also play a role here, see Section \ref{wide_stack}}

\begin{figure*}
\centering
\includegraphics[width=17cm]{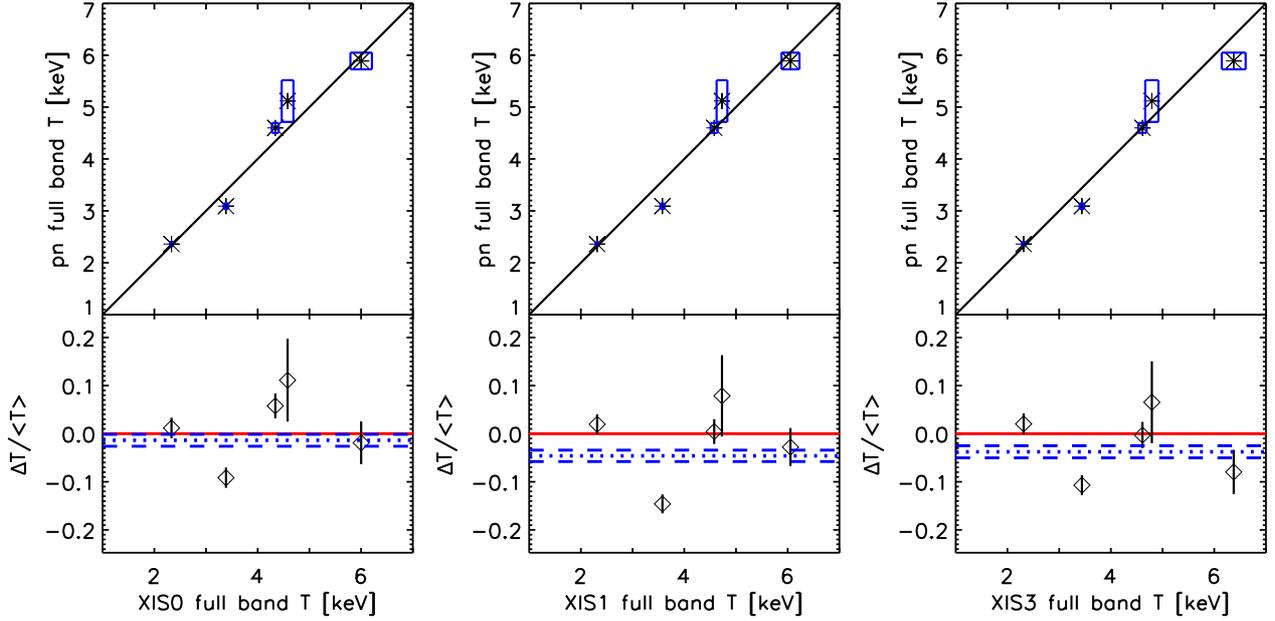}
\vspace{-3.5cm}
\caption{{\it Upper panel:} The best-fit full band temperatures (asterisks) and 1$\sigma$ uncertainties (boxes) for  XMM-Newton EPIC-pn and Suzaku XIS instruments using the public calibration. The solid black line is an identity line, drawn as a reference.
{\it Lower panel:} The relative temperature differences $f_{T}$ (diamonds) between XMM-Newton EPIC-pn and Suzaku XIS instruments, and their 1$\sigma$ uncertainties. The dotted and dashed lines show the weighted means of the relative temperature difference and corresponding uncertainties.
\label{fig:pn_xis_new_fb_public}}
\end{figure*}

Comparison of best-fit temperatures using the full band yielded that the effect of the lowest energy XIS 
cross-calibration uncertainties to the full band temperatures are $\sim$ 3 \%. The effect of Suzaku-XIS / EPIC-pn 
soft energy calibration uncertainties on full band temperatures is $\sim$ 1 -- 5 \%, depending on the XIS instrument 
(see Fig. \ref{fig:pn_xis_new_fb_public}).

\subsubsection{Stacked residuals}
\label{wide_stack}
When using the stacked residuals method for evaluation of the cross-calibration uncertainties of the effective area,
a very detailed physically sound model is not necessary: the usage of the stacked residuals formula (Eq. \ref{R0131}) will take into 
account and correct for the deviations of the best-fit model and the data of the reference instrument. Assuming that the calibration 
of the reference instrument is correct, the phenomenological model (i.e. the best-fit model corrected for the deviations with the data)
derived for the reference instrument still yields a prediction which matches the data of the studied instrument,
if the calibration of the studied instrument is accurate.
 
\begin{figure*}
\centering
\includegraphics[width=17cm]{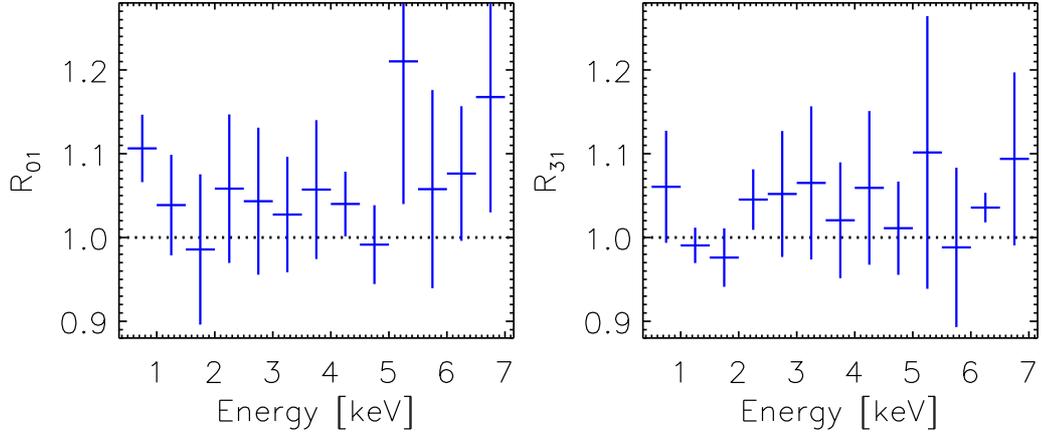}
\caption{The median $\pm$ median absolute deviation of the full energy band stacked residuals of XIS0 / XIS1 R$_{01}$ (left panel) 
and XIS3 / XIS1 R$_{31}$  (right panel) using the public calibration. Note that the cluster sample used here is different from the 
one used for the hard band. The data are regrouped to uniform binning of 0.5 keV.}
\label{fig:xis_mekalmod_rat_0570_sys_med}
\end{figure*}

Yet the stacked residuals for XIS instruments (using XIS1 as a reference as before) in the full band yield
systematic differences at energies above 2 keV (Fig. \ref{fig:xis_mekalmod_rat_0570_sys_med}), 
inconsistent with the above 
argumentation and the fact that in the hard band we found a very good agreement between the different XIS instruments 
(Figs. \ref{fig:new_xis_hb} and \ref{fig:xis_hb_res}). We think that the reason for this is that the sample used for the hard band studies is larger than the one used for 
the soft and full band study due to our selection criteria (Section \ref{sect:sample}). In the hard band the dispersion of the
residuals of individual clusters is bigger for the smaller sample
where one or two deviant objects may cause the difference. 
Thus, the full band stacked residuals analysis above 2 keV energies should be taken with caution.

\begin{figure}
  \resizebox{\hsize}{!}{\includegraphics{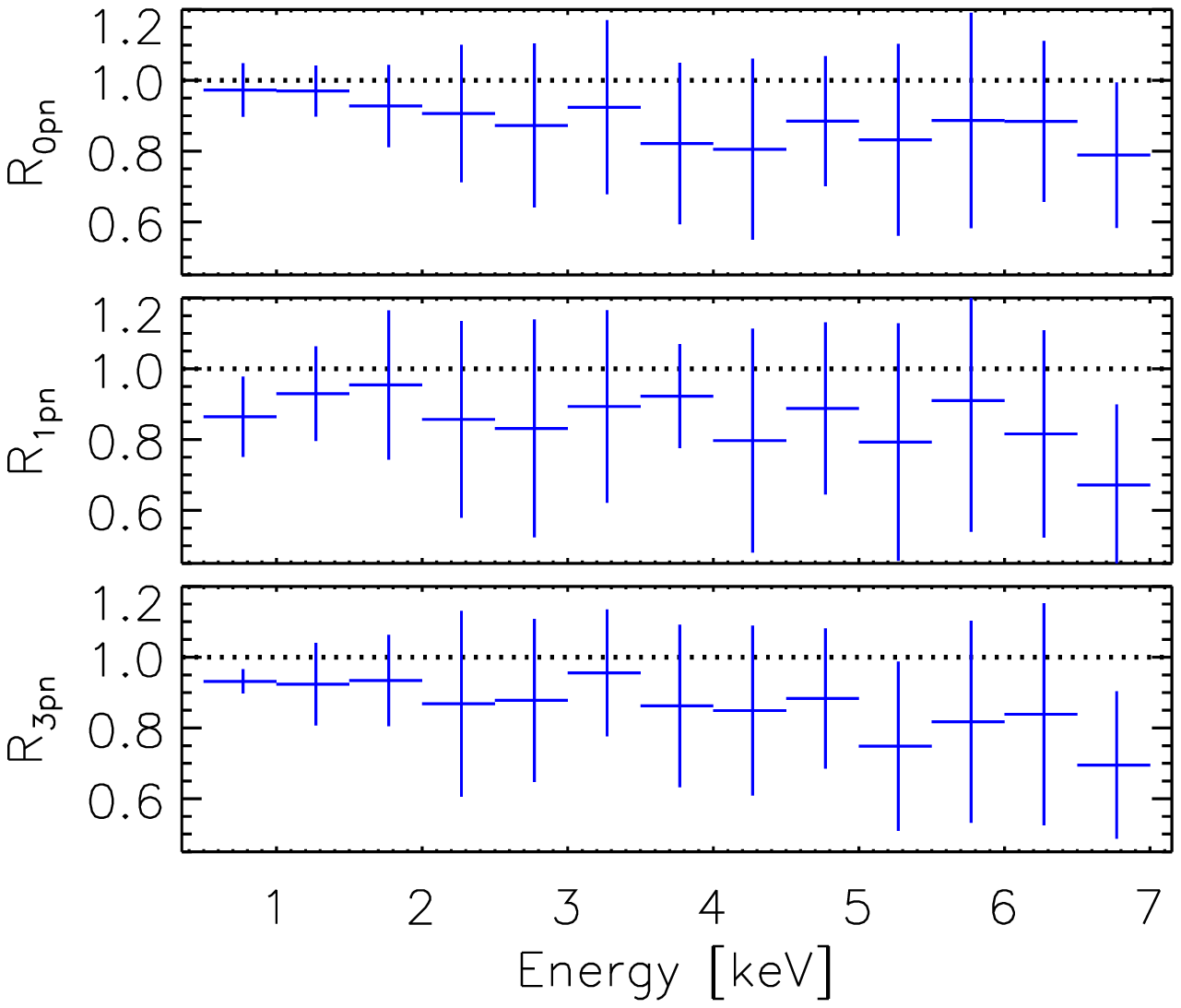}}
    \caption{The median $\pm$ median absolute deviation full band stacked residuals of XI0/pn (R$_{\rm 0pn}$, top panel), 
XIS1/pn (R$_{\rm 1pn}$,  middle panel) and XIS3/pn (R$_{\rm 3pn}$, bottom panel). The data are regrouped to uniform binning of 0.5 keV.
}
    \label{fig:xis_pn_0570_rat}
\end{figure}

A similar problem is present when analysing the stacked residuals of the XIS instruments using EPIC-pn as a reference 
instrument: while the XIS / pn stacked hard band residuals showed a very good consistence 
(Figs. \ref{fig:pn_xis_new_hb} and \ref{fig:xis_pn_2070_rat}), the stacked XIS / pn residuals of the smaller sample used for the full band (Fig. \ref{fig:xis_pn_0570_rat})
exhibit systematic differences at energies above 2 keV. As above, we think that the differences of the samples are causing this discrepancy, and that the XIS / pn cross-calibration uncertainties above 2 keV are better evaluated using the larger hard band sample.
Thus, with the current data we cannot properly estimate the effect of the full band cross-calibration uncertainties 
between the Suzaku-XIS instruments and between the Suzaku-XIS and the EPIC-pn instruments.

\subsection{Suzaku old calibration}
\label{sect:xis_oldnew}

In order to study the effects of the different calibration versions on the discrepancies detected in the XIS data in both hard and 
soft energy bands, we reprocessed the data additionally with the previous Suzaku ftools version 17. The older version (XIS CALDB release 20110210) 
differs mainly by containing the previous model for the XIS contamination (ae\_xi*\_contami\_20081023.fits for XIS1 and XIS3, and ae\_xi*\_contami\_20090128.fits for XIS0) where 
the carbon-to-oxygen ratio was fixed to 6.0.

The different calibration versions changed the best-fit hard energy band temperatures by less than 0.1 \% and the soft energy band temperatures 
by less than 3.5 \%. Thus, temperature discrepancies and consistencies reported above also apply to the old calibration. We will not study the effects 
of different XIS calibration versions in any more detail.

\section{Fluxes}

In addition to the cross-calibration of the shape of the effective area, the cluster data can also be used to assess the 
cross-calibration of the normalisation of the effective area. To the first order the flux of the cluster emission depends linearly
on the accuracy of the normalisation of the effective area. Due to the differences of the composition of the samples used for the 
analysis of the hard and soft band (see Section \ref{sect:sample}), we will only study fluxes in the hard band here.

We examined the hard band fluxes given by the best-fit unabsorbed MEKAL models presented in
(Section \ref{sect:new_xis_hb}). We calculated the fluxes and their statistical uncertainties using the cflux model in Xspec. Due to masked point 
sources, bad columns and pn CCD gaps, the covered fraction of the  3 -- 6 arcmin 
annulus used for spectral extraction varies between the instruments. To enable meaningful comparison of fluxes we calculated the flux in terms of mean 
surface brightness by dividing the total flux from the extraction region with the useful detector area for each cluster. 
Since the emission is not constant with radius, the difference in covered areas might introduce some differences to the flux of a given cluster measured 
with different instruments.

We found that the hard band fluxes measured with XIS instruments agree within a 2 \% statistical uncertainty.
As indicated by the hard band stacked residuals (Fig. \ref{fig:xis_pn_2070_rat}), EPIC-pn fluxes are on average lower than those obtained with 
XIS0, XIS1 and XIS3 instruments, with the difference amounting 
to $\sim$ 10 \% (32.5$\sigma$), $\sim$ 6 \% (20.7$\sigma$) and $\sim$ 6 \% (18.9$\sigma$) respectively (see Table \ref{tab:fluxes}).
For Suzaku-XIS instruments the object-dependent fraction of the PSF scatter of photons from the cluster core to the 3 -- 6 arcmin extraction region might affect the observed fluxes.
For clusters A1795 and A3112 with the brightest cool cores in our sample,
we estimated in Section \ref{sect:PSF} that in maximum $\sim$ 15 \% of the photons in the extraction region may be scattered from the cool core. The fraction is expected to be lower for other clusters in our sample. Thus, XIS fluxes might be biased high 
and a correction for the PSF effect would thus yield the fluxes more consistent with EPIC-pn.

Since the Chandra-ACIS instrument has been reported to yield $\sim$ 10 \% higher fluxes than EPIC-pn in the hard band \citep{nevalainen10},
we find perfect agreement between XIS0 and ACIS fluxes and XIS1 and XIS3 are in closer agreement with ACIS than pn. However, correcting XIS fluxes for 
the maximum PSF scatter of $\sim$ 15 \% would render all XIS instruments to closer agreement with pn than ACIS. Thus the comparison if XIS hard band 
fluxes are consistent with pn or ACIS is inconclusive. We also note that hard band XIS / pn cross-calibration uncertainties (Fig. \ref{fig:xis_pn_2070_rat}) 
yield a maximum systematic uncertainty of $\sim$ 10 \% to the derived fluxes.

\begin{table*}
\caption{Fluxes in the hard energy band given by the unabsorbed MEKAL model. 
}
\label{tab:fluxes} 
\centering
\begin{tabular}{l | c c c c}
\hline\hline

\multicolumn{1}{c}{} & \multicolumn{4}{c}{Hard energy band (2.0 -- 7.0 keV)} \\
\hline
 Name      & XIS0                   & XIS1                   & XIS3                   & pn                    \\
           & flux \tablefootmark{a} & flux \tablefootmark{a} & flux \tablefootmark{a} & flux \tablefootmark{a} \\
\hline 
A1060 &       1.35 [1.33--1.36] & 1.38 [1.36--1.40] & 1.34 [1.32--1.35] & 1.19 [1.16--1.12]  \\ 
A1795 &       0.82 [0.81--0.83] & 0.67 [0.66--0.68] & 0.73 [0.72--0.74] & 1.34 [1.34--1.36]  \\ 
A262  &       0.63 [0.62--0.64] & 0.61 [0.60--0.62] & 0.60 [0.59--0.61] & 0.55 [0.53--0.56]  \\ 
A3112 &       0.46 [0.45--0.47] & 0.45 [0.44--0.46] & 0.46 [0.45--0.47] & 0.43 [0.42--0.43]  \\ 
A496  &       0.40 [0.40--0.41] & 0.26 [0.25--0.26] & 0.32 [0.31--0.32] & 1.47 [1.45--1.48]  \\ 
AWM7  &       2.52 [2.49--2.55] & 2.50 [2.47--2.52] & 2.49 [2.46--2.52] & 2.18 [2.17--2.18]  \\ 
Centaurus &   2.12 [2.09--2.13] & 1.69 [1.67--1.71] & 1.77 [1.75--1.78] & 2.09 [2.08--2.09]  \\ 
Coma &        5.07 [5.06--5.09] & 5.12 [5.10--5.13] & 5.11 [5.10--5.13] & 4.22 [4.20--4.24]  \\ 
Ophiuchus & 14.19 [14.16--14.23] & 14.14 [14.11--14.18] & 13.91 [13.88--13.95] & 11.53 [11.50--11.59] \\ 
Triangulum &  4.00 [3.97--4.01] & 3.88 [3.85--3.89] & 3.93 [3.92--3.96] & 3.45 [3.40--3.51]  \\ 
\hline 
\end{tabular}
\tablefoot{\tablefoottext{a}{The flux is given in units of 10$^{-13}$ ergs/cm$^2$/s/arcmin$^2$ and corresponds to the 
mean surface brightness in the 3 -- 6 arcmin extraction region.}
}
\end{table*}

\section{Suzaku OBF contamination}
\label{sect:obf_contam}
We examined here whether possible uncertainties in the modelling of the Suzaku OBF contaminant could explain the problems in the cross-calibration 
of the effective areas of different XIS instruments we reported above.

It has long been observed that the XIS optical blocking filter (OBF) suffers from contamination on the spacecraft side.\footnote{See e.g. http://heasarc.nasa.gov/docs/suzaku/analysis/abc/}. 
The contaminant is unknown, but is thought to consist of oxygen, hydrogen and carbon, and possibly arises from a DEHP-like material 
that has been created by evaporation from the spacecraft components. The instrumental and temporal variability of the contaminant 
in the public Suzaku calibration (which used contamination file ae\_xi*\_contami\_20091201.fits) is modelled using 
observations of 1E0102-72.3 and RXJ1856.5-3754. The hydrogen-to-carbon number density ratio is assumed to be a constant 157.61, whereas the 
oxygen-to-carbon ratio varies with time as\footnote{See e.g. http://space.mit.edu/XIS/monitor/}
\begin{equation}
\frac{\rm N_O}{\rm N_C} = 0.27602 \times (1 - \exp(-({\rm MJD}-53595.34)/341.65)), 
\label{eq:o-to-c}
\end{equation}
where MJD is the Mean Julian Day of the observation. The spatial variability of the contaminant is measured with monthly integrated observations of 
the sun-lit Earth.
The contaminant is assumed to have the same composition in all parts of the detector, but its column density varies as a function of radius from 
the centre of the detector as
\begin{equation}
{\rm N_C}~(t,r) = \frac{{\rm N_C}~(t,r = 0)}{1 + (r / {\rm A}(t))^{{\rm B}(t)}}, 
\label{eq:contam_tr}
\end{equation} 
where r is the radius in arcmin, ${\rm N_C}~(t, r = 0)$ is the column density of the contaminant at time $t$ in the centre of the detector and A$(t)$ and 
B$(t)$ are time dependent constants.

\begin{figure}
  \resizebox{\hsize}{!}{\includegraphics{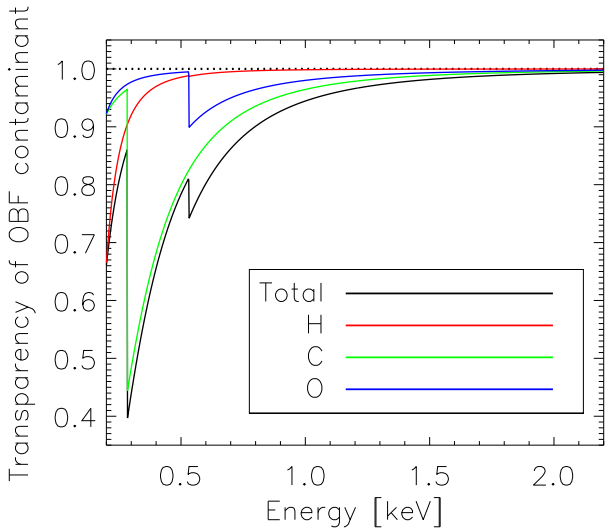}}
    \caption{The total (black) transparency of the OBF contaminant as a function of energy with contributions of individual elements O (blue), C (green) and H (red) for a contaminant with  N$_{\rm O}$ = $0.80 \times 10^{18}$ cm$^{-2}$ and a composition corresponding to an observation date of 2006-10-27 
(approximately coincident with the the mean observation date of our soft band sample), i.e. oxygen-to-carbon ratio of 0.20 and  hydrogen-to-carbon 
ratio of 157.61. 
}
    \label{xis_obf_ratio_plot}
\end{figure}

Since the absorption of the contaminant is strongest in the soft energy band (see Fig. \ref{xis_obf_ratio_plot}), possible uncertainties of the column densities of the oxygen and carbon in the current public calibration of the contaminant may contribute to the soft band uncertainties we reported above (Section \ref{sect:new_xis_sb}).

\begin{figure}
  \resizebox{\hsize}{!}{\includegraphics{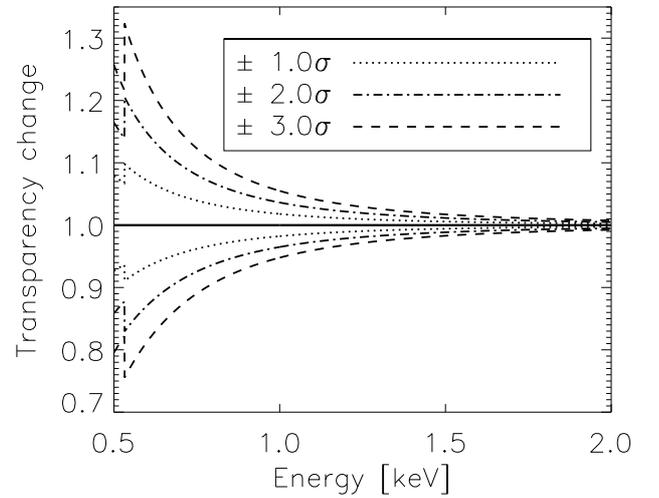}}
    \caption{The change of the transparency of the 2006-10-27 OBF contaminant when the oxygen column density is varied by $\pm$ 1.0$\sigma$ 
(dotted lines), 2.0$\sigma$ (dashed and dotted lines) and 3.0$\sigma$ (dashed lines) while keeping the relative abundance of the other elements fixed.
}
    \label{xis_obf_contam_plot}
\end{figure}

Indeed, if we  adopt an estimate for the systematic uncertainty from on-axis calibration measurements\footnote{from http://space.mit.edu/XIS/monitor/contam/. 
As we use off-axis extraction regions, the uncertainties of the spatial dependence 
of the contaminant will somewhat increase the actual systematic uncertainty for our extraction region.  
The systematic uncertainties of the calibration measurements dominate over the statistical uncertainties.} 
and vary the amount of the 2006-10-27 contaminant (see Fig. \ref{xis_obf_ratio_plot}) by 1 -- 3$\sigma$, 
i.e. vary the oxygen column density by 0.5 -- 1.5 $\times 10^{17}$ cm$^{-2}$ while keeping the relative abundance of other elements constant, 
the effective area changes increasingly with lower energy (see Fig. \ref{xis_obf_contam_plot}). This yields the same 
level of difference ($\sim$ 20 \% at 0.5 keV) as indicated by the Suzaku XIS0 / XIS1 and XIS3 / XIS1 cluster observations (see Fig.   
\ref{fig:xis_mekalmod_rat_simult0520_sys_med}). 
As the change of the oxygen column density does not cause any effect in the hard band, agreement of the XIS instruments with EPIC-pn in the hard band is 
maintained (Section \ref{sect:new_xis_hb}). Thus, a possible uncertainty in the oxygen and carbon column densities is a feasible cause for the 
problems in the cross-calibration of XIS soft band effective areas.

\subsection{Spectral analysis of OBF column densities}
\label{sect:OBF_spect}

\subsubsection{Method}

We proceeded by testing quantitatively the hypothesis that the reported systematic uncertainties of the oxygen column density 
(Section \ref{sect:obf_contam}) are responsible for the soft band XIS calibration problems (Fig. \ref{fig:xis_mekalmod_rat_simult0520_sys_med}). 
The test includes modifying the XIS effective areas with a physically motivated absorption model for the contaminant and requiring that all 
instruments produce an equal emission model for a given cluster.

We adopted the model used by the Suzaku calibration team  as the absorption model, with absorption cross-sections from 
\citet{b-c92}. We determined the 
oxygen-to-carbon ratio for each cluster according to the observation date using the time dependence (Eq. \ref{eq:o-to-c}) and used the 
hydrogen-to-carbon ratio of the contaminant in the public calibration.
\footnote{Since the data do not enable the detection of both the C and O edge, we are not able to test the calibration of the O-to-C ratio} 
By allowing the oxygen column density to vary, we thus varied the total amount of the contaminant, while keeping the composition of the model 
consistent with the contaminant implemented in the public calibration.

We multiplied the MEKAL emission model with the above absorption model, while forcing the temperature and metal abundance equal in all instruments for a given cluster. We allowed the normalisation of the emission model for a given cluster to be an independent parameter for the different instruments. We created auxiliary response files assuming no contaminant, and used these to fit the product of the MEKAL model and 
absorption model. While maintaining the emission models equal between all instruments for a given cluster, the fit to data determines the required 
oxygen column density, i.e. we fit the effective areas simultaneously with the cluster emission.

\subsubsection{Quality of the modelling}
\begin{figure}
  \resizebox{\hsize}{!}{\includegraphics{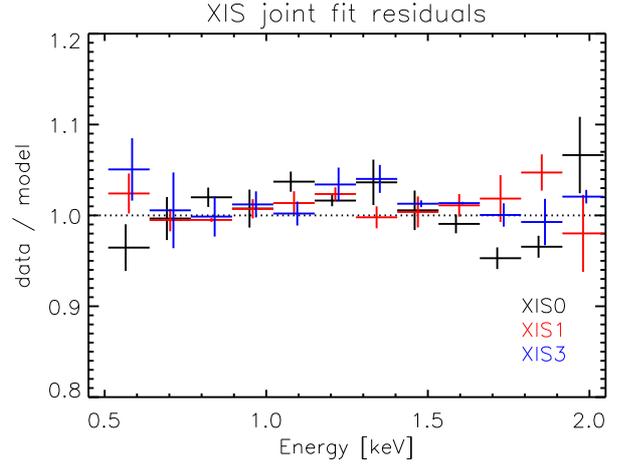}}
    \caption{The median and the median absolute deviation of the residuals (data / best-fit model) of the soft band cluster sample for XIS0 (black), XIS1 (red) 
and XIS3 (blue). The model is an absorbed single temperature MEKAL whose temperatures and metal abundances are forced equal in all instruments for a given 
cluster, fitted to 0.5 -- 2.0 keV band using the effective areas modified with the contaminant model. 
    \label{fig:xis_sb_cont_residuals}}
\end{figure}

By regrouping the spectra to uniform binning of 127.75 eV (see Section \ref{public_t}), we found that the above modification 
to XIS effective areas does in all cases improve the fits over fits using the public calibration 
(Figs. \ref{fig:xis_sb_residuals}  and \ref{fig:xis_sb_cont_residuals}; Tables \ref{tab:simult_par} 
and \ref{tab:fit_par}). The improvement is clearly significant, as is evident from visual inspection of 
the residuals. However, the standard F-test evaluation of the improvement in terms of the decrease of $\chi^2$ relative to 
the decrease of the degrees of freedom is not applicable here. While the $\chi^2$ in the modified ARF fits {\it decrease}, 
there are more degrees of freedom in the modified ARF fits than in the public ARF fits, as the number of free parameters in the public ARF 
fits is 9  (3 temperatures, 
3 abundances and 3 normalisations) while in the modified ARF fits the number of free parameters is 8 (1 temperature, 
1 abundance, 3 normalisations and 3 values of N$_{\rm O}$).
\footnote{as spectral  fit parameters relating to the sky background (see Section \ref{skybkg}) are forced equal in all 
instruments for modified ARF fits, the difference in degrees of freedom increase further}

\subsubsection{Modification to the contaminant}
When we apply the stacked residuals method (Section \ref{effarea}) to the results with the modified effective areas, we find that the median of
the ratios R01 and R31 become close to unity (Fig. \ref{fig:xis_mekalmod_rat_simult0520_sys_med}).  Thus these measures of the effective area 
cross-calibration uncertainties improve over those obtained with the public calibration.  The stacked residuals with modified ARFs are consistent 
with unity in all channels except 0.5 -- 0.6 keV for the XIS3 / XIS1 ratio.
This implies that while the current public H--C--O composition with 
modified column density yields a significant improvement over the column density implemented in the current public calibration, there may still be room 
for adjusting the composition of the contaminant and other constituents of the effective area.

In order to evaluate the cluster-based modification to the column density of the contaminant implemented in the current calibration, we need to work 
out the values of the oxygen column density in our 3--6 arcmin extraction regions in the public calibration, since these are not available in the 
auxiliary response files produced by the standard software.

Currently the radial behaviour of the contaminant (Eq. \ref{eq:contam_tr}), essential for extended sources like clusters, is only measured for the 
XIS1 instrument and assumed equal for XIS0 and XIS3. This is a possible source of remaining calibration uncertainties which we cannot examine with 
the current data. 
We divided our 3 -- 6 arcmin extraction regions 
into concentric annuli with a width of 0.5 arcmin and calculated the N$_{\rm O}$ of the public calibration at the centres of these annuli.
We weighted these values with the flux in each annulus using the surface brightness profiles of  \citet{nevalainen10} for A1795, A262 and A3112; 
\citet{markevitch99} for A496 and \citet{tamura00} for A1060, in order to recover the N$_{\rm O}$ values in the current calibration. The flux in the 
annuli remains rather constant with radius due to the opposing effects of radially decreasing surface brightness and radially increasing area of the annuli.
Thus, the flux-weighted average oxygen column densities are within $\pm$ 1 \% of that at a distance of 4.5 arcmin from the centre of the FOV and we used this
to evaluate the oxygen column densities of the contaminant for our clusters in the public calibration.

\begin{figure*}
\sidecaption
  \includegraphics[width=12cm]{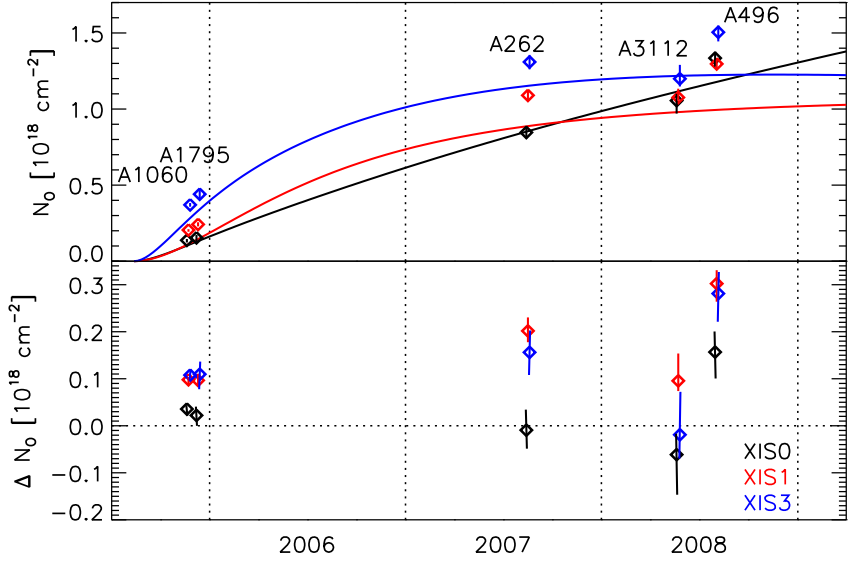}
\caption{The upper panel shows the time dependence of the oxygen column densities and 1.0 $\sigma$ statistical uncertainties of the contaminant fitted to the cluster data (diamonds) and the ones implemented in the public calibration at 4.5 arcmin distance from the centre of the FOV (lines) for XIS0 (black), XIS1 (red) and XIS3 (blue). The lower panel 
shows the time dependence of the difference of these. 
\label{fig:xis_contam_time}}
\end{figure*}

Most of the modified oxygen column densities (see Fig.\ref{fig:xis_contam_time} and Table \ref{tab:nO}) are within 
3$\sigma$ of the reported uncertainties of the values implemented in the public calibration (see Section 
\ref{sect:obf_contam}). In more detail, the cluster analysis indicated that the N$_{\rm O}$ values of XIS0 are very close 
to public calibration values up to May 2008 while XIS1 and XIS3 require additional oxygen column density of the contaminant up to 
2 $\times 10^{17}$ cm$^{-2}$. This implies that the soft band effective area shape of XIS0 is currently accurately 
calibrated and that the effective area of XIS1 and XIS3  is overestimated increasingly towards the lowest energies in 
the public calibration for observations up to to May 2008.

All instruments require a significant increase of the oxygen column density for A496 observation on Aug 2008. This effect 
is not evident in the monitoring data of E0102-72.
\footnote{http://www.astro.isas.ac.jp/suzaku/doc/suzaku\_td/node10.html\#contami}
This may be affected by the fact that the Galactic N$_{\rm H}$ varies by$\sim 2 \times 10^{20}$ cm$^{-2}$ within 1$^{\circ}$ from the 
cluster centre \citep{kalberla05}.  Also, allowing the column density to be a free parameter, \citet{tanaka06} derived
a N$_{\rm H}$ profile for A496 which increases towards the cluster centre, reaching a maximum of  $\sim 6 \times 10^{20}$ cm$^{-2}$ in 
the 3 -- 6 arcmin annulus, while we used the LAB average of $\sim 4 \times 10^{20}$ cm$^{-2}$ in our fits. 
From consideration of the transmission curves (see Fig. \ref{xis_obf_ratio_plot}), we conclude that this discrepancy in the Galactic 
N$_{\rm H}$ can account for the extra contaminant required in the fits to A496.

\begin{table*}
\caption{The oxygen column density of the 0.5--2.0 keV band spectral fits when forcing the temperatures (kT) and metal abundances (abund) equal in all 
instruments for a given cluster and allowing the oxygen column density (N$_{\rm O}$) of the contaminant to be a free parameter for all sources and instruments. 
$\Delta$ N$_{\rm O}$ describes the modification to the oxygen column density of the contaminant. The uncertainties 
are the statistical ones given at 1$\sigma$ level.
}
\label{tab:nO} 
\centering
\begin{tabular}{l|cc|cc|cc}
\hline\hline
         & \multicolumn{2}{c|}{XIS0}      & \multicolumn{2}{c|}{XIS1} & \multicolumn{2}{c}{XIS3} \\  
 Name   & N$_{\rm O}$  & $\Delta$ N$_{\rm O}$  & N$_{\rm O}$ & $\Delta$ N$_{\rm O}$ &   N$_{\rm O}$ & $\Delta$ N$_{\rm O}$   \\
        & [$10^{18}$ cm$^{-2}$] & [$10^{18}$ cm$^{-2}$] & [$10^{18}$ cm$^{-2}$] & [$10^{18}$ cm$^{-2}$]& [$10^{18}$ cm$^{-2}$] & [$10^{18}$ cm$^{-2}$]   \\  
\hline 
A1060   & 0.14 [0.12--0.15] & 0.04 [0.02 -- 0.04]    & 0.20 [0.19--0.21] & 0.10 [0.09 -- 0.10] & 0.37 [0.36--0.38] & 0.11 [0.10 -- 0.12] \\ 
A1795   & 0.15 [0.13--0.17] & 0.02 [0.00 -- 0.04]    & 0.24 [0.23--0.26] & 0.10 [0.08 -- 0.11] & 0.44 [0.41--0.47] & 0.11 [0.08 -- 0.14] \\
A262    & 0.85 [0.81--0.89] & 0.00 [-0.05 -- 0.03]   & 1.09 [1.07--1.12] & 0.20 [0.18 -- 0.23] & 1.31 [1.26--1.36] & 0.16 [0.11 -- 0.20] \\
A3112   & 1.06 [0.97--1.10] & -0.06 [-0.15 -- -0.02] & 1.08 [1.05--1.13] & 0.10 [0.07 -- 0.15] & 1.20 [1.15--1.29] & -0.02 [-0.07 -- 0.07] \\
A496    & 1.33 [1.28--1.38] & 0.16 [0.10 -- 0.20]    & 1.30 [1.26--1.33] & 0.30 [0.26 -- 0.33] & 1.50 [1.44--1.55] & 0.28 [0.22 -- 0.33] \\
\hline 
\end{tabular}
\end{table*}

\subsubsection{New cluster emission models}

\begin{figure*}
\centering
\includegraphics[width=17cm]{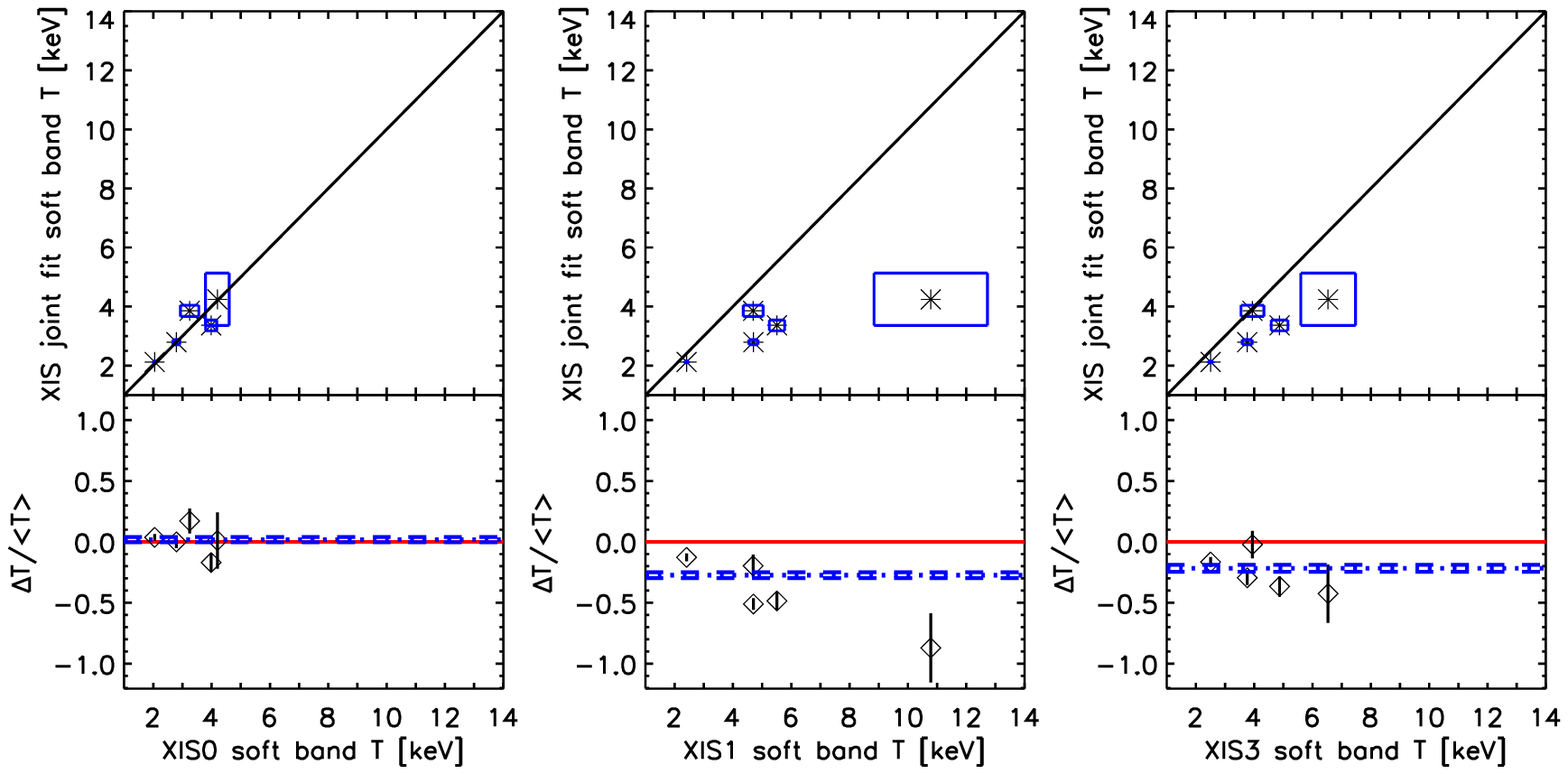}
\vspace{-3.5cm}
\caption{
{\it Upper panel:} The best-fit soft band temperatures (asterisks) and 1$\sigma$ uncertainties (boxes) for Suzaku XIS instruments using the modified 
XIS effective area (joint fit) and public calibration. The solid black line is an identity line, drawn as a reference.
{\it Lower panel:} The relative temperature differences $f_{T}$ (diamonds) between the Suzaku XIS instruments, and their 1$\sigma$ uncertainties. The dotted and dashed lines show the weighted means of the relative temperature difference and corresponding uncertainties.
}
\label{fig:xis_joint_publ}
\end{figure*}

Since the required modification of the XIS0 effective area yields only minor changes for the oxygen column density of the contaminant 
(see Fig. \ref{fig:xis_contam_time}), the corresponding changes on the XIS0 temperatures are small (see Fig. \ref{fig:xis_joint_publ}, 
and Tables \ref{tab:simult_par} and  \ref{tab:fit_par}).
The required higher N$_{\rm O}$ for XIS1 and XIS3 increases the absorption and in order to produce the same model prediction, 
the emission models become softer. On average the XIS1 and XIS3 temperatures decrease by $\sim$ 27 \% (11.5$\sigma$) and $\sim$ 22 \% (7.4$\sigma$) 
respectively when compared to the fit with the public calibration.

\begin{figure}
  \resizebox{\hsize}{!}{\includegraphics{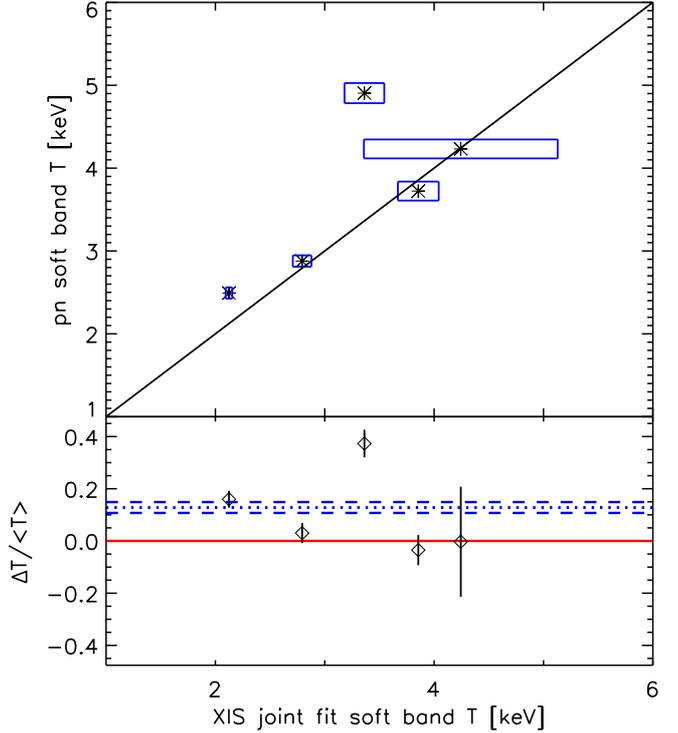}}
    \caption{{\it Upper panel:} The best-fit soft band temperatures (asterisks) and 1$\sigma$ uncertainties (boxes) for EPIC-pn and Suzaku XIS instruments 
using the modified XIS effective area. The solid black line is an identity line, drawn as a reference.
{\it Lower panel:} The relative temperature differences $f_{T}$ (diamonds) between EPIC-pn and Suzaku, and their 1$\sigma$ uncertainties. The dotted and dashed 
lines show the weighted means of the relative temperature difference and corresponding uncertainties.
\label{fig:pn_XIS_newcal_nocontam_soft_band}}
\end{figure}

Since the soft band temperatures of XIS1 and XIS3 decrease significantly with the modification of the contaminant, the Suzaku-XIS v.s. EPIC-pn soft band temperature comparison will change from that using the public calibration, presented in Section \ref{public_t}. The joint XIS0 + XIS1 + XIS3 fit to the soft band with the free contaminant 
yielded on average $\sim$ 12 \% (6.2$\sigma$) lower temperatures compared to EPIC-pn soft bands temperatures (see Fig. 
\ref{fig:pn_XIS_newcal_nocontam_soft_band}). 
As noted in Section \ref{sect:PSF}, the PSF scattering from cool cores may bias XIS best-fit temperatures low by 7 \% in the 
worst case; however, we expect this bias to be much smaller in most cases. If corrected for the PSF scatter, the 
XIS soft band temperatures might increase by a few \%, i.e., become closer to the EPIC-pn values.
Since the Chandra-ACIS instrument has been found to yield $\sim$18\% higher temperatures than EPIC-pn in the soft band \citep{nevalainen10}, our results indicate that the XIS instruments yield a better soft band temperature agreement with the EPIC-pn instrument than with the Chandra-ACIS instrument.
The data exhibit quite a lot of scatter: three of the clusters yield consistent values between Suzaku-XIS and EPIC-pn, while the temperatures of two of 
the clusters differ by 4.9$\sigma$ and 7.1$\sigma$ respectively.

\begin{figure}
  \resizebox{\hsize}{!}{\includegraphics{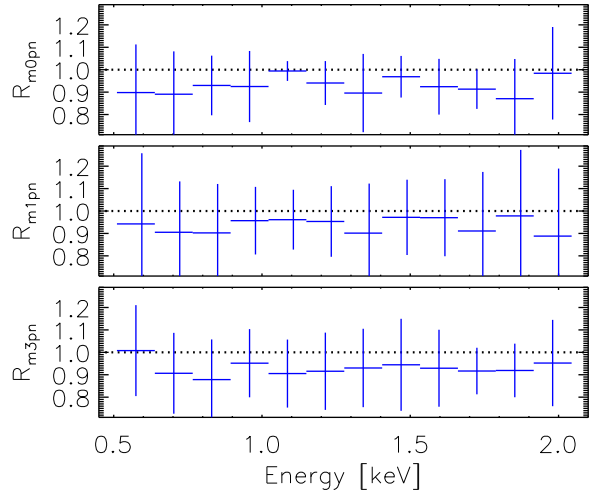}}
    \caption{The median $\pm$ median absolute deviation of the soft energy band stacked residuals of XIS0 / pn R$_{\rm m0pn}$ (top panel), XIS1 / pn 
R$_{\rm m1pn}$ (middle panel) and XIS3 / pn R$_{\rm m3pn}$ (bottom panel) using the  contamination-modified effectve areas for XIS instruments. The data are regrouped to uniform binning of 127.75 keV.}
    \label{fig:pn_xis_mod_soft_stacked}
\end{figure}

With the introduction of the modified contaminate the Suzaku-XIS1,3 / EPIC-pn stacked soft band residuals at the lowest energies become consistent with 
unity, while XIS0 and pn remain consistent in both cases (compare Figs. \ref{fig:pn_xis_mekalmod_rat_2070_sys_med}
and \ref{fig:pn_xis_mod_soft_stacked}). However, the scatter is larger (10 -- 20 \%) when using the modified contaminate, and thus
the remaining energy-dependent uncertainties, which amount to 12 \% temperature differences (see above), are not significantly detected with the current data.

\begin{table}
\caption{The temperatures, metal abundances and $\chi^2 / dof$ of the 0.5--2.0 keV band spectral fits when forcing the temperatures (kT) and abundances (abund) 
equal in all instruments for a given cluster and allowing the oxygen column density (N$_{O}$) of the contaminant to be a free parameter for all sources and 
instruments. The uncertainties are the statistical ones given at 1$\sigma$ level.
}
\label{tab:simult_par} 
\centering
\begin{tabular}{l|ccc}
\hline\hline
 Name   & kT               & abund              & $\chi^2 / dof$ \\ 
        & [keV]            & Solar              &                \\ 
\hline 
A1060   & 2.8 [2.7--2.9]   & 0.37 [0.35--0.39]  & 748.18/613     \\ 
A1795   & 4.2 [3.0--4.8]   & 0.16 [0.12--0.24]  & 242.37/246     \\ 
A262    & 2.1 [2.1--2.2]   & 0.55 [0.52--0.59]  & 444.91/380     \\ 
A3112   & 3.9 [3.7--4.1]   & 0.33 [0.29--0.37]  & 334.84/326     \\ 
A496    & 3.4 [3.2--3.6]   & 0.29 [0.26--0.36]  & 506.83/501     \\ 
\hline 
\end{tabular}
\end{table}

\section{Conclusions}

\begin{table}
\caption{The relative difference ($\mu$) and significance (sig.) of the temperatures and fluxes obtained with different instrument combinations. }
\label{tab:summary} 
\centering
\begin{tabular}{l|c c |c c |c c}
\hline\hline
Instruments &\multicolumn{2}{c|}{T$_{\rm hard}$ \tablefootmark{a}} & \multicolumn{2}{c|}{T$_{\rm soft}$ \tablefootmark{b}} 
& \multicolumn{2}{c|}{flux$_{\rm hard}$ \tablefootmark{a}}\\ 
            & $\mu$ \tablefootmark{c} & sig. \tablefootmark{d}    & $\mu$ \tablefootmark{c} & sig. \tablefootmark{d}  & $\mu$ \tablefootmark{c} & sig. \\
\hline 
ACIS / pn \tablefootmark{e}   & -1   & 0.6 &  18 &  8.6  & 11 & 24.7 \\
pn / XIS0                     &  5   & 4.2 &  14 &  6.2  & -5 & 1.1  \\
pn / XIS1                     &  2   & 1.7 & -18 &  7.4  & -1 & 0.2  \\
pn / XIS3                     &  6   & 4.9 &  -9 &  3.0  & -1 & 0.2  \\
pn / XIS013 \tablefootmark{f} &  --  & --  &  12 &  6.2  & -- & --   \\
XIS0 / XIS1                   & -5   & 5.6 & -29 & 11.2  & 4  & 1.0  \\
XIS0 / XIS3                   & -1   & 0.7 & -23 &  7.4  & 4  & 1.0  \\
XIS1 / XIS3                   &  5   & 4.9 &   9 &  2.9  & 0  & 0.0  \\
\hline 
\end{tabular} 
\tablefoot{The values correspond to the best-fit single temperature models in \tablefoottext{a}{2.0 -- 7.0 keV} and
\tablefoottext{b}{0.5 -- 2.0 keV} energy bands using the public calibration, i.e. for Suzaku this refers 
XIS CALDB version 20110608 and XRT CALDB version 20080709 or using the modification to the XIS contaminant \tablefoottext{f}{} 
as explained in the text. \tablefoottext{c}{$\mu$ gives the weighted mean of the relative difference between 
the measured temperatures in percentages}; \tablefoottext{d}{sig. gives the statistical significance of the temperature difference 
in terms of $\sigma$} ; \tablefoottext{e}{from \citet{nevalainen10}}.
}
\end{table}

\begin{figure*}
\sidecaption
  \includegraphics[width=12cm]{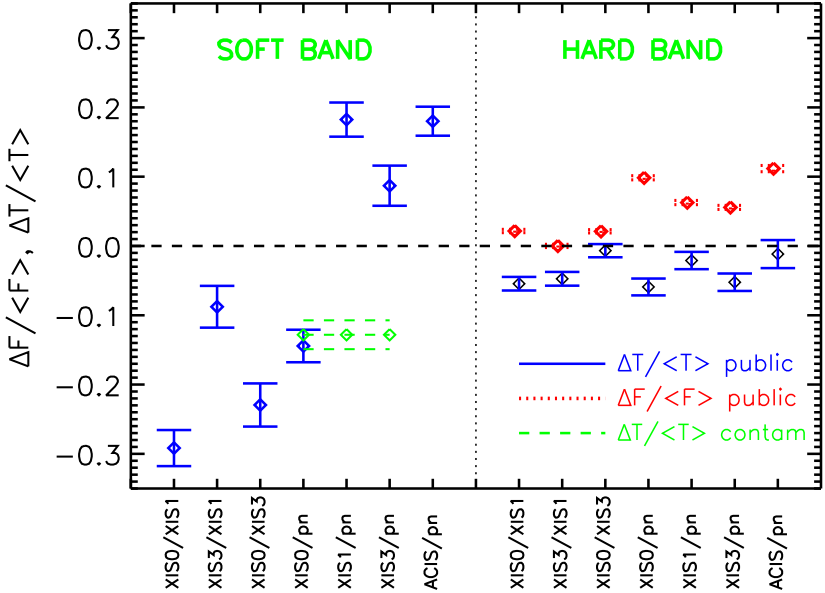}
\caption{The average relative difference (diamonds) $\pm$ the error of the mean of the temperatures (solid line) 
and fluxes (dotted line) using the public calibration in the soft band (left side of the plot) and in the hard band 
(right side of the plot). Comparison of the pn and XIS soft band temperature using the modification to the contaminant
is shown with diamonds and dashed line.}
\label{fig:t_flux_plot}
\end{figure*}

We used a sample of galaxy clusters to perform a study on the accuracy of the cross-calibration of the energy dependence {\textbf and the normalisation} of the effective 
area of XIS0, XIS1 and XIS3 instruments onboard Suzaku and EPIC-pn instrument onboard XMM-Newton in 3 -- 6 arcmin annular extraction regions. 
The Suzaku results apply to the XIS CALDB versions 20110608 (i.e. public calibration at the writing of this paper) and 20110201, and to XRT 
CALDB version 20080709 for observations carried out in the time period of years 2005 -- 2008.
The XMM-Newton results refer to SAS version xmmsas\_20110223\_1801-11.0.0 and the latest calibration information in June 2011.
{\textbf The results are summarised in Fig. \ref{fig:t_flux_plot} and Table \ref{tab:summary}.}

As the shape of the spectrum determining the temperature of the cluster depends on the {\textbf energy dependence} effective area, 
{\textbf and the flux depends on the normalisation of the effective area}
we performed the study by comparing the temperatures, fluxes and best-fit emission models of the clusters measured with different instruments in the 0.5 -- 2.0 and 2.0 -- 7.0 keV bands. 

Comparison of the Suzaku-XIS instruments with each other yielded that XIS0 and XIS3 produced rather consistent hard band temperatures which are systematically lower than those obtained with XIS1 by $\sim$ 5 \%.
The Suzaku-XIS hard band temperatures are systematically lower than those obtained with EPIC-pn instrument of XMM-Newton, 
but only by 2 -- 6 \% in average.
Some fraction of this difference could be caused by Suzaku's PSF scattering from the cool core, which should be less than $\sim$ 6 \% in the worst case.
Since EPIC-pn hard band temperatures are consistent with Chandra-ACIS and BeppoSAX-MECS temperatures, and EPIC-pn
bremsstrahlung temperatures agree with Fe XXV/XXVI line ratio temperatures \citep{nevalainen10}, we suggest that 
the energy dependence of XIS instruments is quite accurately calibrated in the hard band. The remaining calibration uncertainties may 
yield effects on the XIS hard band temperatures at $\sim$ 6 \% level.

In the soft 0.5 -- 2.0 keV energy band the XIS1 and XIS3 instruments yielded consistent temperatures when using the public calibration, while 
XIS0 yielded temperatures which are lower by an average of 29 \% (11.2$\sigma$)  and 23 \% (7.4$\sigma$) from those obtained with XIS1 and XIS3, respectively. 
Comparison of the residuals showed that XIS0 effective area is underestimated or XIS1 and XIS3 effective area is overestimated towards lower energies, with the 
difference increasing to $\sim$ 20 \% at 0.5 keV. 

We tested the assumption that the above discrepancies in the effective area in the soft band are due to uncertainties of the implemented 
column density of the optical blocking filter (OBF) contaminant. We achieved this by allowing the OBF contaminant to have a 
column density different from that in the public calibration and fitting jointly the XIS0 + XIS1 + XIS3 soft band spectral 
data, while forcing the temperatures and metal abundances equal in all instruments for a 
given cluster. The resulting fits to the data were significantly better than the ones using the public calibration.
Assuming the composition of the contaminant as in the public calibration, we found that the cluster data required an 
addition to the column density of the contaminant of XIS1 
and XIS3 amounting to a maximum $\Delta$N$_{\rm O}$ of 2 $\times 10^{17}$ cm$^{-2}$, i.e. at a 4$\sigma$ level of the 
reported uncertainties of the N$_{\rm O}$ measurements in the public calibration. The column density of the XIS0 contaminant 
implemented in the public calibration is consistent with that obtained in our spectral analysis of clusters.
Thus, the analysis implies that the effective area of XIS0 in the soft band is more accurately modelled in the public 
calibration than that of XIS1 and XIS3, for observations during 2005 -- 2008.

The XIS soft band temperatures obtained with the modified effective area are lower by $\sim$ 12 \% from those obtained 
with EPIC-pn instrument onboard XMM-Newton. Considering that the PSF scatter from the cool cores may bias XIS soft band 
temperatures low, the PSF-corrected XIS soft band temperatures would become in better agreement with those obtained using EPIC-pn. However, Chandra-ACIS instrument yields $\sim$ 18 \% higher temperatures than EPIC-pn in the soft band (\citep{nevalainen10}). 
Thus, our results indicate that the XIS instruments yield a better soft band temperature agreement with the EPIC-pn instrument than with the Chandra-ACIS instrument.

The hard band fluxes derived with Suzaku-XIS instruments are higher 
than those derived with EPIC-pn by $\sim$ 6 -- 10 \%, whereas the Chandra-ACIS instrument yields $\sim$ 10 \% higher fluxes than EPIC-pn in the hard band 
\citep{nevalainen10}. However, as Suzaku-XIS fluxes might be biased high by a maximum of $\sim$ 15 \% due to PSF scatter and Suzaku-XIS and EPIC-pn hard 
band data shows that the derived fluxes have a maximum systematic uncertainty of $\sim$ 10 \%, comparison of Suzaku-XIS to EPIC-pn and Chandra-ACIS fluxes 
proved inconclusive.

\acknowledgements
We would like to acknowledge IACHEC members and thank L. David, K. Hamaguchi, H. Matsumoto, M. Nobukawa and M. St\"uhlinger
for useful comments. KK acknowledges support from Magnus Ehrnrooth foundation and 
Nylands Nation. EDM acknowledges support from NASA grant NNX09AE58G to MIT. This research has made use of data obtained 
from the High Energy Astrophysics Science Archive Research Center (HEASARC), provided by NASA's
Goddard Space Flight Center.

\bibliographystyle{aa} 
\bibliography{bibliography}

\appendix

\section{Spectral fits}
\label{sect:spect_fits}

\begin{figure*}
\centering
\includegraphics[width=17cm]{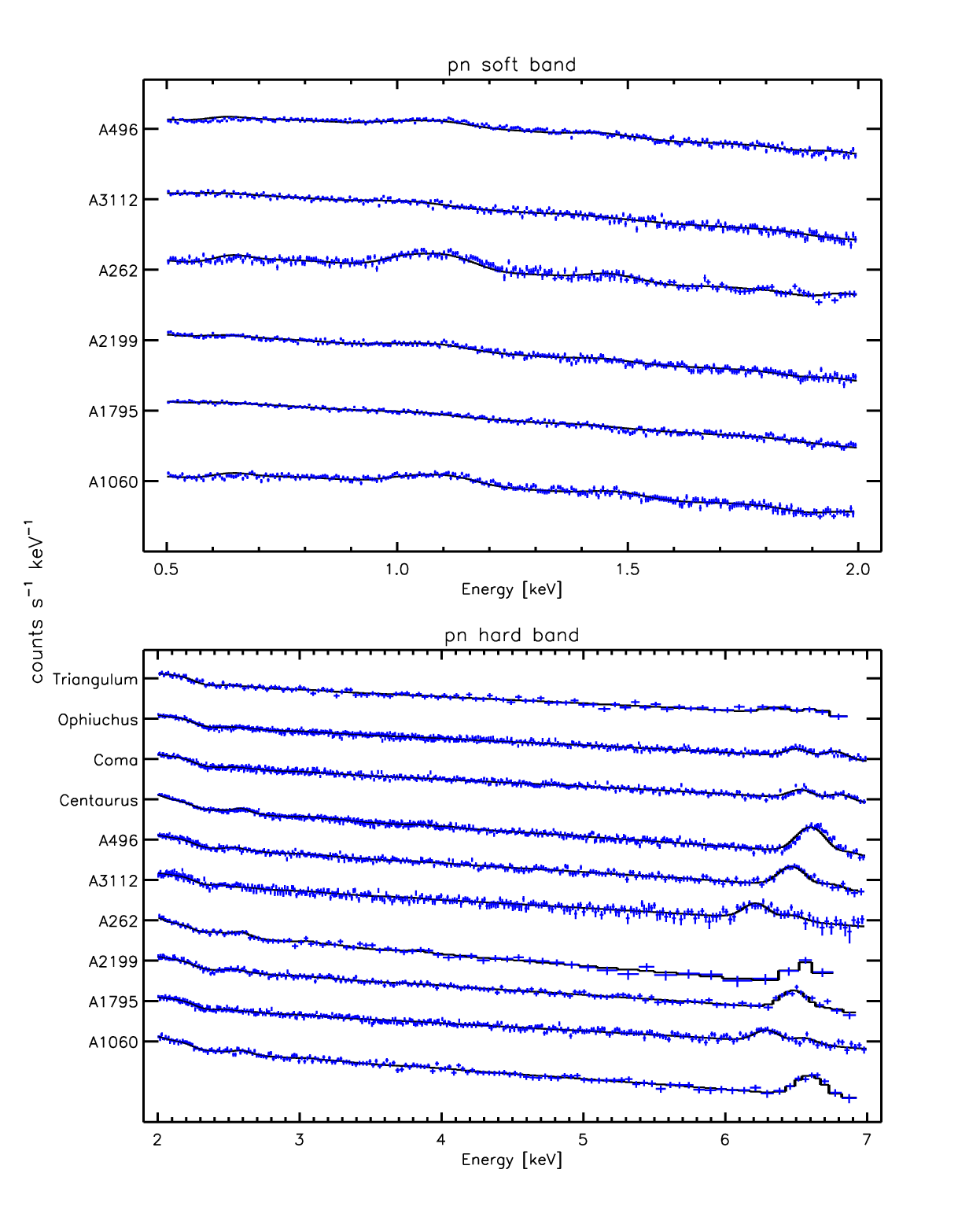}
\caption{The soft band (upper panel) and hard band (lower panel) XMM-Newton EPIC / pn spectra (crosses) and best-fit single temperature independent 
fits (solid lines) for the 
cluster sample. The normalisations of the spectra are adjusted for plot clarity and do not reflect the relative brightness of the clusters.
}
\label{fig:pn_spec}
\end{figure*}

\begin{figure*}
\centering
\includegraphics[width=17cm]{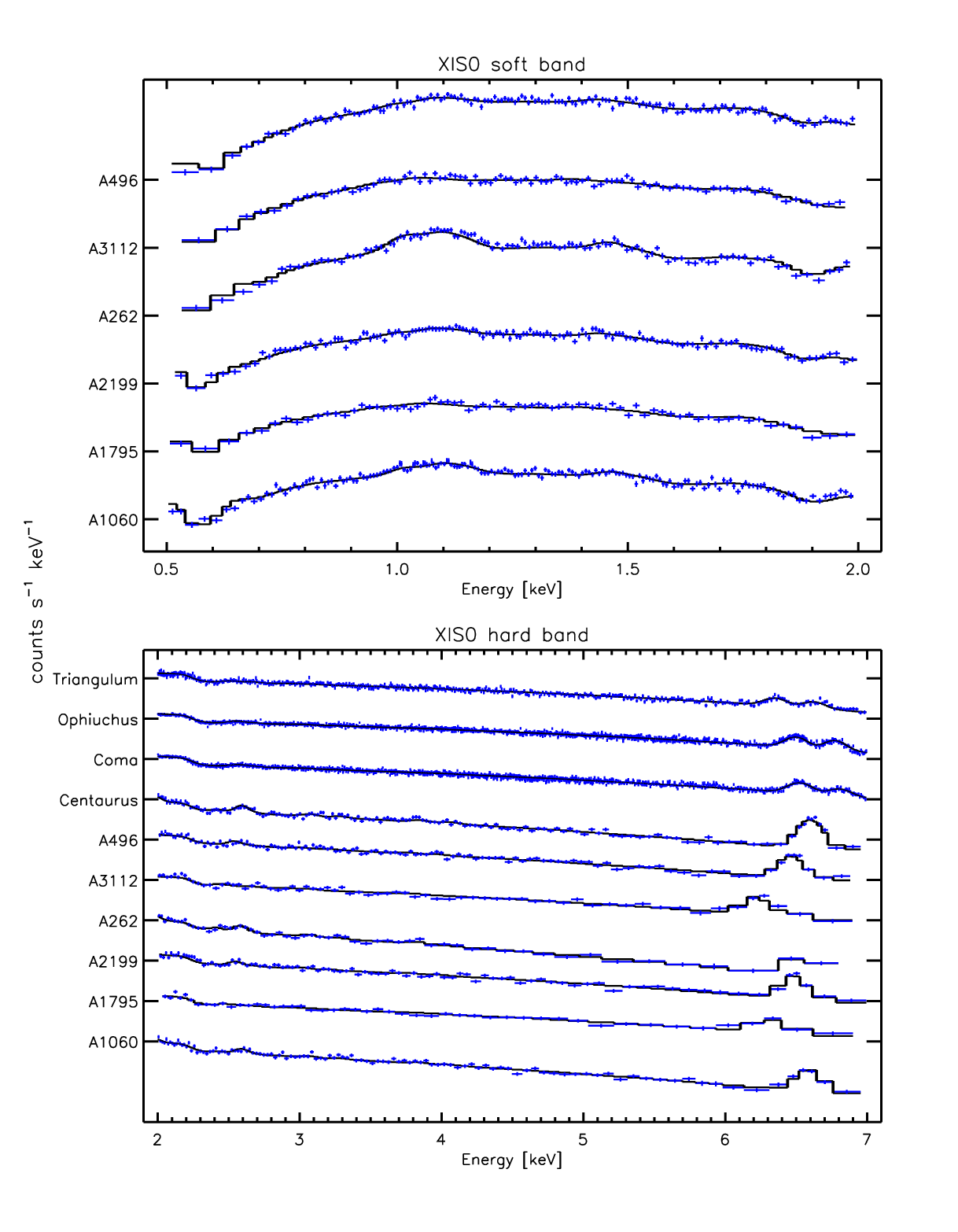}
\caption{The soft band (upper panel) and hard band (lower panel) Suzaku XIS0 spectra (crosses) and best-fit single temperature independent fits 
(solid lines) for the cluster sample. The normalisations of the spectra are adjusted for plot clarity and do not reflect the relative brightness 
of the clusters.
}
\label{fig:xis0_spec}
\end{figure*}

\begin{figure*}
\centering
\includegraphics[width=17cm]{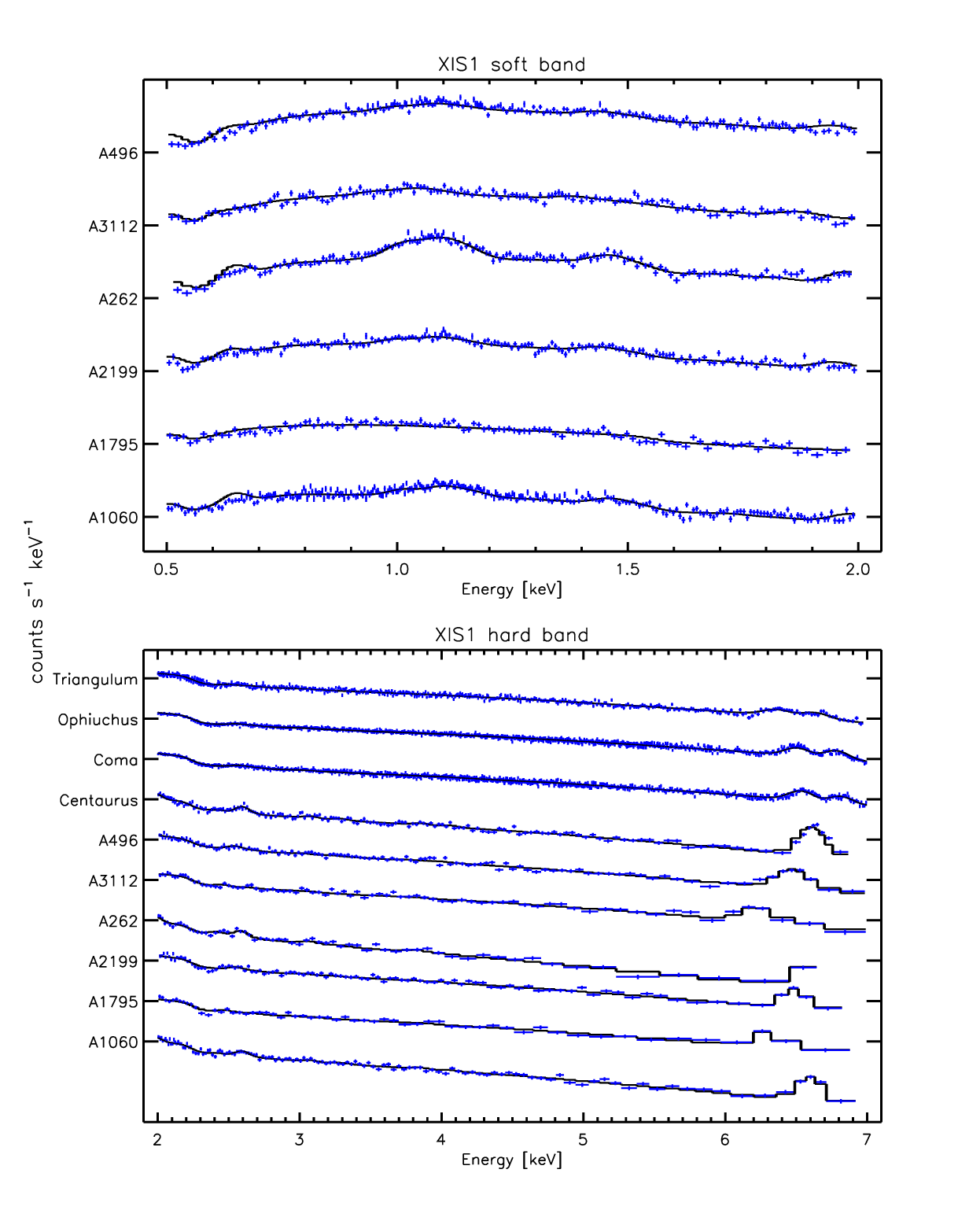}
\caption{The soft band (upper panel) and hard band (lower panel) Suzaku XIS1 spectra (crosses) and best-fit single temperature independent fits 
(solid lines) for the cluster sample. The normalisations of the spectra are adjusted for plot clarity and do not reflect the relative brightness 
of the clusters.
}
\label{fig:xis1_spec}
\end{figure*}

\begin{figure*}
\centering
\includegraphics[width=17cm]{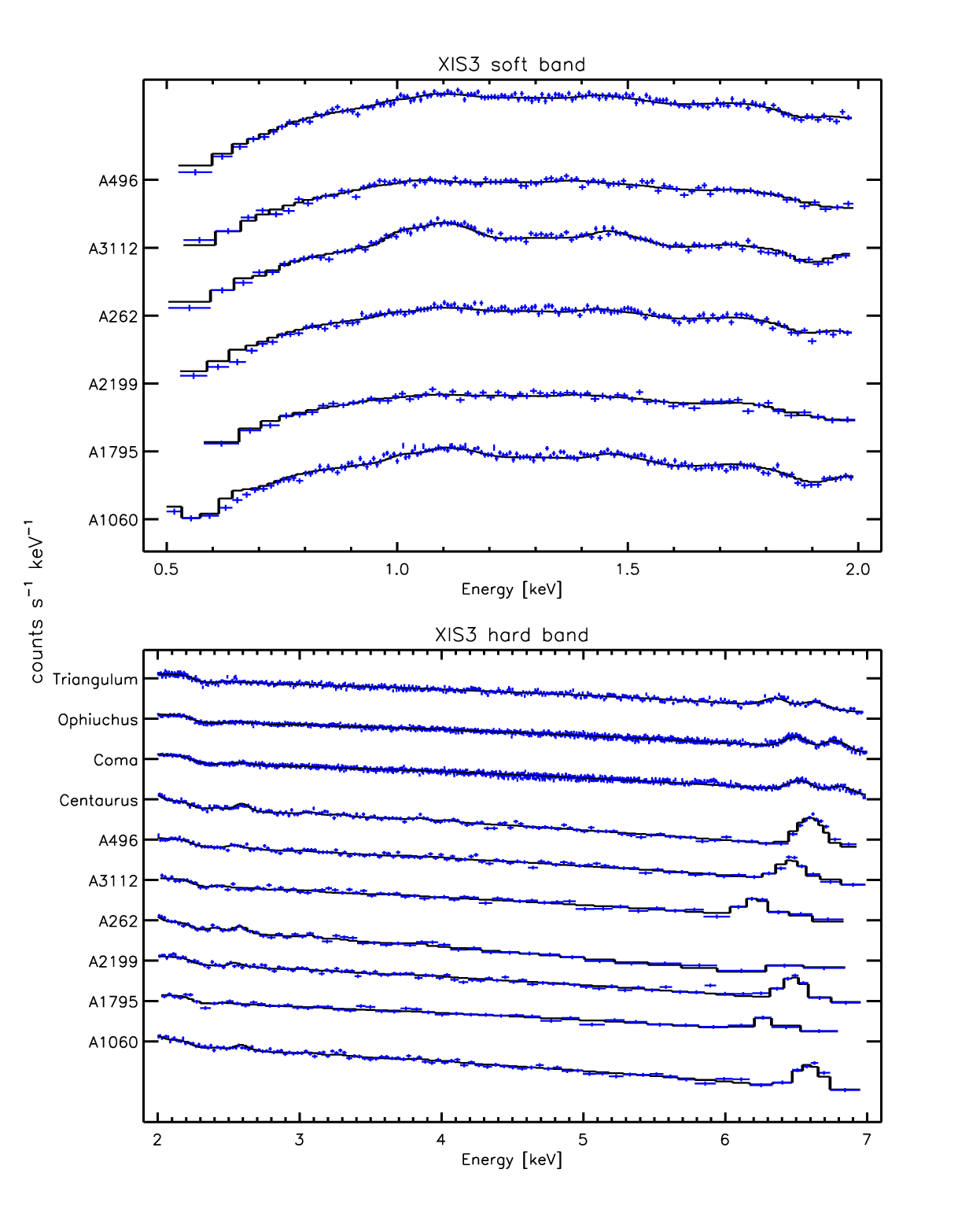}
\caption{The soft band (upper panel) and hard band (lower panel) Suzaku XIS3 spectra (crosses) and best-fit single temperature independent fits 
(solid lines) for the cluster sample. The normalisations of the spectra are adjusted for plot clarity and do not reflect the relative brightness 
of the clusters.
}
\label{fig:xis3_spec}
\end{figure*}

Here we show the data and the best-fit single temperature MEKAL models for EPIC-pn (Fig. \ref{fig:pn_spec}), XIS0 (Fig. \ref{fig:xis0_spec}), 
 XIS1 (Fig. \ref{fig:xis1_spec}) and XIS3 (Fig. \ref{fig:xis3_spec}) using the public calibration. The spectral parameters are listed in Table \ref{tab:fit_par}.

\begin{table*}
\caption{Best-fit parameters of the Suzaku XIS and XMM-Newton EPIC-pn spectral fits using the public calibration}
\label{tab:fit_par} 
\centering 
\begin{tabular}{l | c c c | c c c } 
\hline\hline 
            & \multicolumn{3}{c|}{XIS0}    & \multicolumn{3}{c}{XIS1}       \\
Name        & kT   & Abund & $\chi^2/dof$ &  kT   & Abund & $\chi^2/dof$     \\
            & [keV] & [Solar] &           & [keV] & [Solar]    &            \\ 
\hline
\multicolumn{1}{c}{} & \multicolumn{6}{c}{Soft energy band (0.5 -- 2.0 keV)} \\
\hline
A1060       & 2.8 [2.7--2.9] & 0.43 [0.39--0.48] & 229.51/170 & 4.7 [4.6--4.9] & 0.98 [0.91--1.05] & 413.27/275 \\ 
A1795       & 4.2 [3.8--4.6] & 0.26 [0.15--0.35] & 73.02/66   &11.0 [9.7--13.5]& 0.03 [0.00--0.32] & 125.64/108 \\
A262        & 2.0 [2.0--2.1] & 0.51 [0.47--0.62] & 166.66/108 & 2.4 [2.3--2.5] & 0.85 [0.80--0.91] & 215.40/159 \\
A3112       & 3.2 [3.0--3.6] & 0.21 [0.14--0.29] & 84.00/86   & 4.7 [4.3--5.0] & 0.61 [0.49--0.69] & 165.84/140 \\
A496        & 4.0 [3.8--4.2] & 0.45 [0.37--0.55] & 170.79/146 & 5.5 [5.3--5.8] & 0.85 [0.75--0.95] & 289.28/205 \\
\hline 
\multicolumn{1}{c}{} &  \multicolumn{6}{c}{Hard energy band (2.0 -- 7.0 keV)} \\
\hline
A1060       & 3.5 [3.4--3.6] & 0.44 [0.40--0.48] & 109.55/106  & 3.4 [3.3--3.5] & 0.53 [0.48--0.58] & 135.32/121 \\
A1795       & 6.6 [6.2--7.0] & 0.40 [0.34--0.47] & 43.04/41    & 5.9 [5.5--6.2] & 0.39 [0.33--0.46] & 56.73/51   \\
A262        & 2.5 [2.5--2.6] & 0.45 [0.40--0.52] & 43.91/50    & 2.5 [2.4--2.5] & 0.52 [0.45--0.60] & 76.81/57   \\
A3112       & 5.4 [4.7--5.7] & 0.45 [0.40--0.51] & 78.00/60    & 5.5 [5.3--5.9] & 0.40 [0.35--0.46] & 69.98/74   \\
A496        & 4.2 [4.1--4.3] & 0.49 [0.45--0.53] & 149.27/119  & 4.4 [4.3--4.6] & 0.51 [0.47--0.56] & 130.02/129 \\
AWM7        & 3.9 [3.8--4.0] & 0.58 [0.53--0.63] & 118.18/97   & 4.0 [3.9--4.1] & 0.71 [0.66--0.77] & 112.21/105 \\
Centaurus   & 3.6 [3.6--3.7] & 0.81 [0.77--0.86] & 171.03/159  & 3.8 [3.7--3.9] & 0.82 [0.77--0.87] & 223.60/177 \\
Coma        & 8.3 [8.2--8.4] & 0.36 [0.35--0.38] &1152.77/1124 & 9.0 [8.8--9.1] & 0.37 [0.36--0.38] &1155.40/1104\\
Ophiuchus   & 9.9 [9.8--10.0]& 0.47 [0.46--0.49] &1337.33/1242 &10.7 [10.5--10.8]& 0.46 [0.44--0.47]&1282.61/1187\\
Triangulum  & 9.5 [9.3--9.8] & 0.31 [0.29--0.33] & 512.69/537  &10.3 [10.0--10.6]& 0.35 [0.32--0.38]& 547.64/536 \\
\hline\hline 
            & \multicolumn{3}{c|}{XIS3}    & \multicolumn{3}{c}{pn}         \\
Name        & kT   & Abund & $\chi^2/dof$ &  kT   & Abund & $\chi^2/dof$     \\
            & [keV] & [Solar] &           & [keV] & [Solar]    &            \\ 
\hline
\multicolumn{1}{c}{} & \multicolumn{6}{c}{Soft energy band (0.5 -- 2.0 keV)} \\
\hline
A1060       & 3.8 [3.5--3.9] & 0.76 [0.68--0.83] & 239.18/159 & 2.9 [2.8--2.9] & 0.48 [0.45--0.50] & 327.00/290 \\
A1795       & 6.5 [5.6--7.5] & 0.51 [0.28--0.82] & 61.45/61   & 4.2 [4.1--4.3] & 0.20 [0.18--0.23] & 331.04/296 \\
A262        & 2.5 [2.4--2.6] & 0.75 [0.67--0.83] & 102.89/102 & 2.5 [2.4--2.5] & 0.71 [0.67--0.78] & 313.94/240 \\
A3112       & 3.9 [3.7--4.5] & 0.32 [0.24--0.45] & 90.14/89   & 3.7 [3.6--3.8] & 0.24 [0.22--0.28] & 353.08/296 \\
A496        & 4.9 [4.6--5.1] & 0.65 [0.53--0.76] & 131.56/141 & 4.9 [4.8--5.0] & 0.72 [0.69--0.78] & 472.49/297 \\
\hline 
\multicolumn{1}{c}{} & \multicolumn{6}{c}{Hard energy band (2.0 -- 7.0 keV)} \\
\hline
A1060       & 3.4 [3.3--3.4] & 0.49 [0.44--0.53] & 108.91/104 & 3.5 [3.3--3.6] & 0.42 [0.37--0.47] & 124.03/167 \\
A1795       & 6.6 [6.2--7.1] & 0.34 [0.28--0.40] & 50.94/42   & 7.3 [7.1--7.5] & 0.35 [0.32--0.38] & 324.32/400 \\
A262        & 2.4 [2.3--2.4] & 0.41 [0.36--0.48] & 46.48/48   & 2.5 [2.4--2.6] & 0.50 [0.41--0.60] & 38.66/78   \\
A3112       & 5.3 [5.0--5.5] & 0.45 [0.39--0.50] & 77.74/63   & 5.9 [5.6--6.2] & 0.42 [0.38--0.46] & 311.65/334 \\
A496        & 4.6 [4.2--4.8] & 0.43 [0.39--0.47] & 122.60/114 & 5.0 [4.9--5.1] & 0.46 [0.43--0.49] & 277.85/353 \\
AWM7        & 4.1 [3.9--4.2] & 0.59 [0.55--0.64] & 109.10/96  & 4.0 [4.0--4.0] & 0.57 [0.56--0.59] & 883.38/806 \\
Centaurus   & 3.6 [3.5--3.7] & 0.81 [0.77--0.86] & 176.12/157 & 3.8 [3.8--3.9] & 0.71 [0.69--0.73] & 814.58/655 \\
Coma        & 8.6 [8.5--8.7] & 0.35 [0.34--0.36] &1356.87/1151& 9.0 [8.8--9.3] & 0.35 [0.33--0.38] & 505.40/531 \\
Ophiuchus   & 9.9 [9.7--10.0]& 0.47 [0.46--0.48] &1392.54/1257&10.4 [10.2--10.7]&0.41 [0.38--0.43] & 677.17/703 \\
Triangulum  & 9.8 [9.6--10.1]& 0.32 [0.30--0.35] & 573.44/524 &11.3 [10.3--12.7]&0.40 [0.29--0.50] & 62.88/106  \\
\hline
\end{tabular}
\end{table*}

\end{document}